\documentclass[aps,prx,twocolumn,groupedaddress]{revtex4-2}
\usepackage{times}
\usepackage{amsmath}
\usepackage{graphicx}
\usepackage{color}
\usepackage{comment}
\usepackage{bm}
\usepackage{amssymb}
\usepackage{amsthm}
\usepackage{bm}
\usepackage{braket}
\usepackage{graphicx}
\usepackage{microtype}
\usepackage{algorithm}
\usepackage{times}
\usepackage{algpseudocode}
\usepackage[breaklinks]{hyperref}
\hypersetup{colorlinks=true, linkcolor=blue, citecolor=blue, filecolor=blue, urlcolor=blue}
\newcommand{\REVISE}[1]{{#1}}

\begin{document}

\title{Hyper-optimized \REVISE{approximate} contraction of tensor networks with arbitrary geometry}

\author{Johnnie Gray}
\author{Garnet Kin-Lic Chan}
\affiliation{Division of Chemistry and Chemical Engineering,  California Institute of Technology, Pasadena, USA 91125}

\begin{abstract}
Tensor network contraction is central
to problems ranging from many-body physics to computer science.
We describe how to approximate tensor network contraction through bond compression on arbitrary graphs. In particular, we introduce a
hyper-optimization over the compression and contraction strategy itself to minimize error and cost. We demonstrate that our protocol outperforms both hand-crafted contraction strategies \REVISE{in the literature} as well as recently proposed general contraction algorithms on a variety of synthetic \REVISE{and physical} problems on regular lattices and random regular graphs. We further showcase the power of the approach by demonstrating approximate contraction of tensor networks for frustrated three-dimensional lattice partition functions, dimer counting on random regular graphs, and to access the hardness transition of random tensor network models, in graphs with many thousands of tensors.
\end{abstract}
\maketitle

\section{Introduction}

\begin{figure}[t!]
    \centering
    \includegraphics[width=1.0\linewidth]{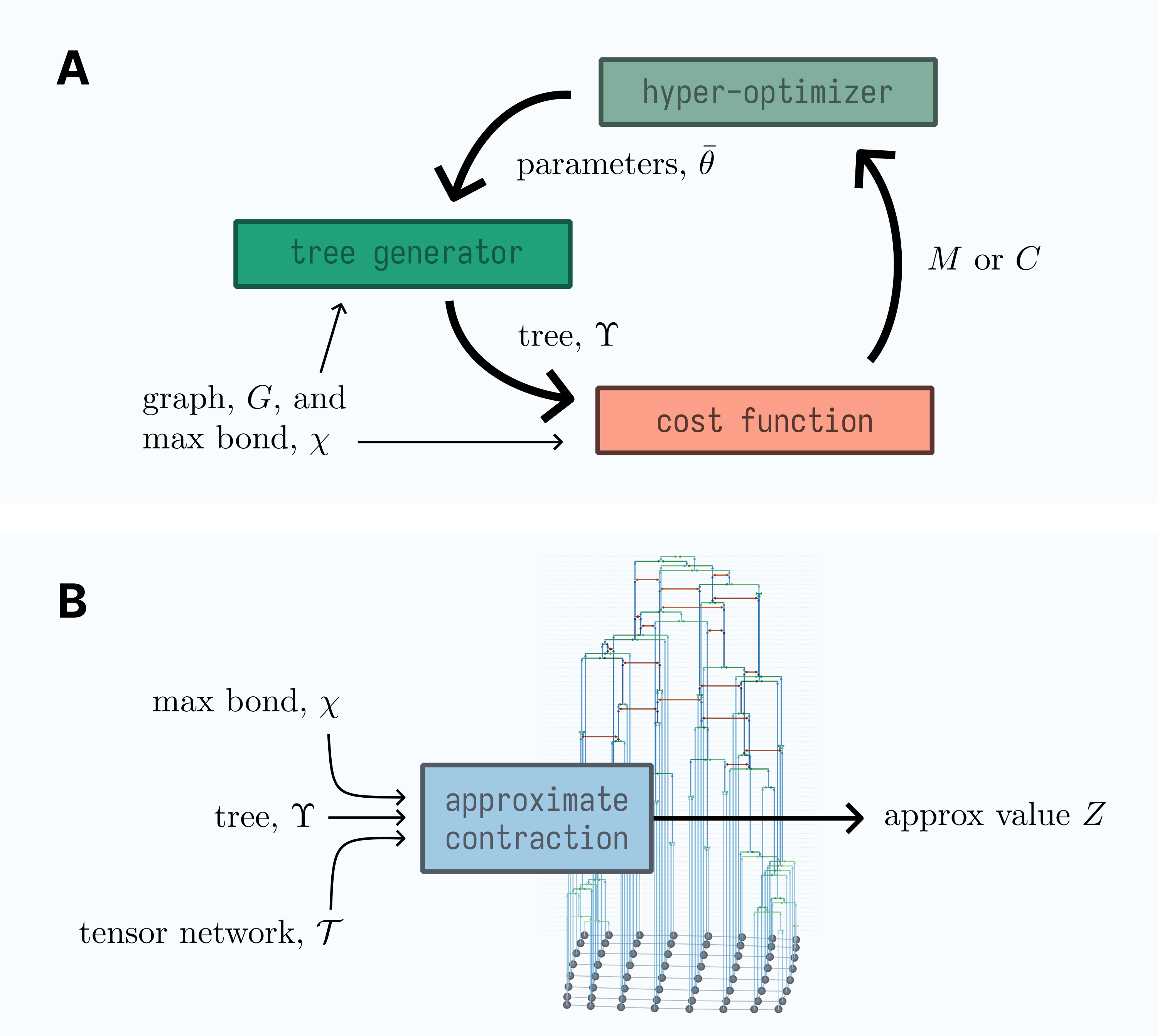}
    \REVISE{\caption{
    Overview. The approximate  contraction is specified by a sequence of contractions and compressions, expressed as an ordered tree. The strategy optimizes a cost-function over such trees.
    \textbf{A}: The hyper-optimization loop. Approximate contraction trees $\Upsilon$ on the graph $G$ are suggested by the tree generator. The tree characteristics are controlled by heuristic parameters $\theta$ and maximum bond dimension $\chi$. The hyper-optimizer minimizes a cost-function $M$ or $C$ (peak memory or computational cost).
    \textbf{B}: The approximate contraction tree. The tensor network $\mathcal{T}$ is shown at the bottom. Moving upwards, pairs of tensors are contracted (blue lines), and singular value compressions are performed between tensors (orange lines). By the top of the tree, one obtains a scalar output $Z$, using resources $\sim M$ or $C$.}
    }\label{fig:hyper-compressed-overview}
\end{figure}

Tensor network contraction, a summation over a product of multi-dimensional quantities, is a common computational structure. For example, this computation underlies quantum circuit simulation
~\cite{markovSimulatingQuantumComputation2008,
pednaultBreaking49QubitBarrier2017,
boixoSimulationLowdepthQuantum2017,
chenClassicalSimulationIntermediateSize2018,
villalongaFlexibleHighperformanceSimulator2018,
grayHyperoptimizedTensorNetwork2021,
huangClassicalSimulationQuantum2020,
kalachevRecursiveMultiTensorContraction2021,
panSimulatingSycamoreQuantum2021},
quantum many-body simulations
~\cite{whiteDensitymatrixAlgorithmsQuantum1993,
whiteDensityMatrixFormulation1992,
murgVariationalStudyHardcore2007,
VerstraeteRenormalizationalgorithmsQuantumMany2004,
vidalClassicalSimulationInfiniteSize2007,
jiangAccurateDeterminationTensor2008,
stoudenmireStudyingTwoDimensionalSystems2012,
VlaarSimulationthreedimensionalquantum2021},
evaluating classical partition functions
~\cite{nishinoCornerTransferMatrix1996,
nishinoCornerTransferMatrix1997,
levin2007tensor,
orusSimulationTwoDimensional2009,
xieSecondRenormalizationTensorNetwork2009,
VanderstraetenResidualentropiesthreedimensional2018,
zhaoTensorNetworkAlgorithm2015,
liu2021tropical},
decoding quantum error correcting codes
~\cite{ferris2014tensor,
bravyi2014efficient,
tuckett2019tailoring,
chubb2021statistical,
chubb2021general,
bonilla2021xzzx,
farrellyTensorNetworkCodes2021},
counting solutions of satisfiability problems
~\cite{schuch2007computational,
garcia2011exact,
biamonteTensorNetworkContractions2015,
kourtis2019fast,
dudek2019efficient,
dudek2020parallel,
de2020tensor,
liu2021tropical},
statistical encoding of natural language
~\cite{gallego2017language,
pestun2017tensor,
bradley2020modeling,
meichanetzidis2020grammar},
and many other applications.
The cost of exact contraction scales, in general, exponentially with the number of tensors.
However, there is evidence, for example in some many-body physics applications, that tensor networks of interest can often be approximately contracted with satisfactory and controllable accuracy,  without necessarily incurring exponential cost~\cite{verstraete2008matrix,orus2019tensor}.
Many different approximation strategies for tensor network contraction have been proposed
~\cite{VerstraeteRenormalizationalgorithmsQuantumMany2004,
murgVariationalStudyHardcore2007,
levin2007tensor,
dittrichCoarseGrainingMethods2012,
evenblyTensorNetworkRenormalization2015,
yangLoopOptimizationTensor2017,
balRenormalizationGroupFlows2017,
VanderstraetenResidualentropiesthreedimensional2018,
VlaarSimulationthreedimensionalquantum2021,
ranLectureNotesTensor2020}.
\REVISE{Especially in many-body physics contexts, the approximate contraction algorithms are usually tied to the geometry of a structured lattice.
In this work, we consider how to search for an optimal approximate tensor network contraction strategy, within an approach that can be used not only for structured lattices, but also for arbitrary graphs. We view the essential prescription
as the order in which contractions and approximate compressions are performed:
this sequence can be summarized as a computational tree with contraction and tensor bond compression steps. Within this framework, sketched in Fig.~\ref{fig:hyper-compressed-overview}, the problem reduces to optimizing a cost function over such computational trees: we term the macro-optimization over trees ``hyper-optimization''.
As we will demonstrate in several examples, optimizing a simple cost function related to the memory or computational cost of the contraction also leads to an approximate contraction tree with small contraction error.
Consequently, our hyper-optimized approximate contraction enables the efficient and accurate simulation of a wide range of graphs encountered in different tasks, bringing the possibility of eliminating, or otherwise improving on, formal exponential costs. In addition, in the structured lattices arising in many-body physics simulations, we observe that we can  improve on the best physically motivated approximate contraction schemes in the literature.}

\begin{figure*}[t!]
\includegraphics[width=1.0\textwidth]{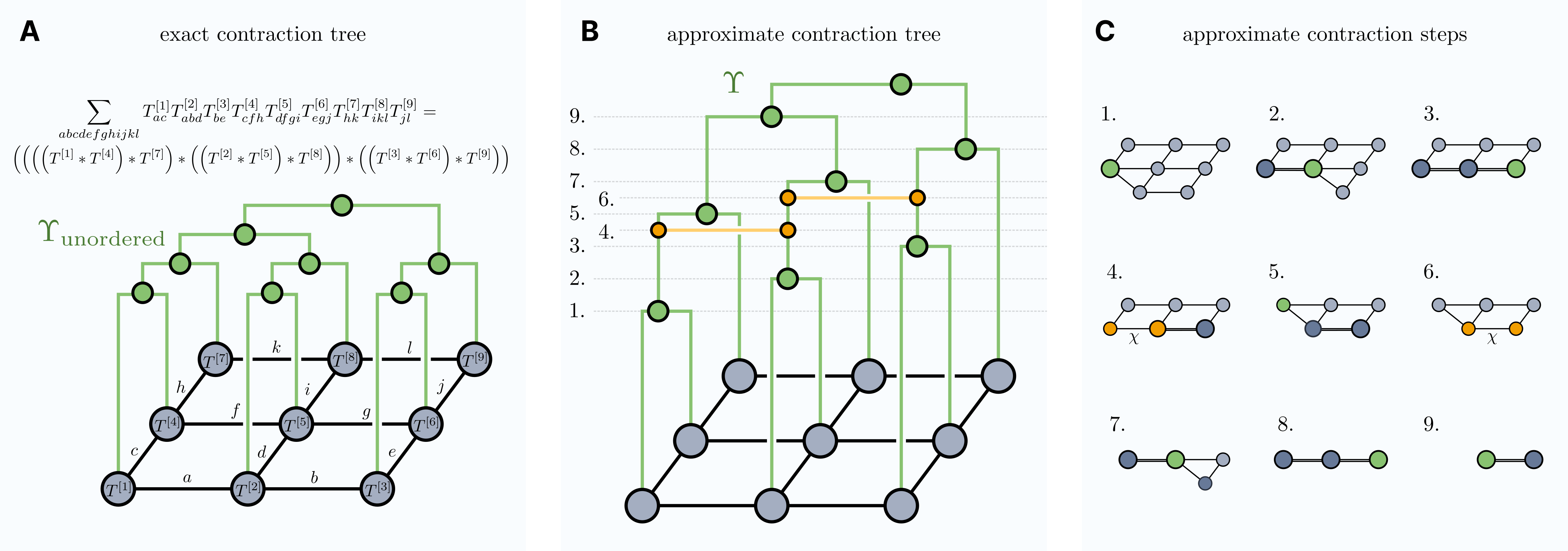}
\caption{
  \REVISE{
  \textbf{A}. Pairwise exact contraction of a tensor network, with the unordered contraction tree $\Upsilon_\text{unordered}$ indicating the contractions.
  Each intermediate (green node) corresponds to a pair of parentheses in the expression.
  \textbf{B}. An approximate contraction tree $\Upsilon$ for same network. Since compression steps do not commute, this tree is ordered. Here, the compressions (orange lines) take place at steps 4.~and 6.
  \textbf{C}. The sequence of contractions and compressions associated with the tree in B. Newly contracted tensors in green, tensors with compressed bonds in orange.
  }
}\label{fig:tndiagram}
\end{figure*}

A tensor $T$ is a multi-index quantity (i.e.~a multi-dimensional array). We use lower indices to index into the tensor, e.g. $T_{i_1i_2 \ldots i_n}$ is an element of an $n$-index $T$, and upper indices to label a specific tensor, e.g. $T^{[1]}, T^{[2]} \ldots$, out of a set of tensors. A tensor network contraction sums (contracts) over the (possibly shared) indices of a product of tensors,
\begin{align}
T_{\{ e_\mathrm{out} \}} = \sum_{ \{ e\} / \{ e_\mathrm{out} \} }  \prod_v T^{[v]}_{\{ e_v \}} \label{eq:tncontract}
\end{align}
where $\{ e\}$ is the total set of indices, $\{ e_\text{out}\}$ is the subset that is left uncontracted, and $\{ e_v\}$ is the subset of $\{ e\}$ for tensor $T^{[v]}$.
We can place the tensors at the vertices $v$ of a network (graph), with the bonds (edges) corresponding to the indices $\{ e\}$. An examples of contraction is shown in  Fig.~\ref{fig:tndiagram}A.

In practice the sum in Eq.~\ref{eq:tncontract} is performed as sequence of pairwise contractions, and the order of contraction
greatly affects both the memory and time costs. Much recent work has been devoted to optimizing  contraction paths in the context of simulating quantum circuits~\cite{grayHyperoptimizedTensorNetwork2021,huangClassicalSimulationQuantum2020,kalachevRecursiveMultiTensorContraction2021,panSimulatingSycamoreQuantum2021}. Parametrized heuristics that efficiently sample the space of contraction paths, for example by graph-partitioning, are crucial, and optimizing the parameters of such heuristics (hyper-optimization) to minimize the overall cost has proven particularly powerful, leading to dramatic reductions in contraction cost (i.e.~many orders of magnitude).

Here we extend the ideas of hyper-optimized tensor network contraction to the setting of approximate tensor network contraction. As discussed above, approximate contraction has a long history in many-body simulation, but such work has focused on regular lattices. Although several recent contributions have addressed arbitrary graphs~\cite{jermyn2020automatic,pan2020contracting,chubb2021general},
with a fixed contraction strategy, they do not focus on optimizing the strategy itself.
In part, this is because there is a great deal of flexibility (and thus many components to optimize) when formulating an approximate contraction algorithm, and because an easily computable metric of quality is not clear a priori.

\REVISE{We proceed by first formulating the search space of approximate tensor network contraction algorithms, which we identify as a search over approximate contraction trees. To reduce the search space, we define simple procedures for gauging and when to compress bonds in the tree.
We then discuss how to sample the large space of trees, by optimizing the hyperparameters of a contraction tree generator, with respect to the peak memory or computational cost.
We use numerical experiments to establish the success of the strategy, comparing to existing algorithms  designed for structured lattices and  for arbitrary graphs.
Finally, using the hyper-optimized approximate contraction algorithms, we showcase the range of computational problems that can be addressed, in many-body physics, computer science, and complexity theory, illustrating the power of approximate tensor network computation.}

\section{A Framework for Approximate Contraction Algorithms}

\subsection{Components of approximate contraction}

\REVISE{In an exact tensor network contraction, the computational graph, specified by the sequence of pairs of tensors which are contracted, can be illustrated as a computational contraction tree.}
This is illustrated in Fig.~\ref{fig:tndiagram}A, where the \REVISE{tensor network is shown by the black lattice at the bottom, and the contractions between pairs occur at the green dots in tree},  $\Upsilon_\text{unordered}$. Note that the value and cost of the exact tensor network contraction does not depend on the order in which the \REVISE{contractions} are performed~\footnote{There is a minor effect on memory.}, thus the contraction tree is unordered.
The problem of optimizing the cost of exact contraction is thus a search over contraction trees to optimize the floating point and/or memory costs.

\REVISE{In the process of contracting tensors, one generally creates larger tensors, which share more bonds with their neighbors.} In approximate contraction we aim to reduce the cost of exact contraction by introducing an approximation error. The most commonly employed approximation is to compress the large tensors into smaller tensors (with fewer or smaller indices); this is the type of numerical approximation that we also consider here. The simplest notion of compression arises in matrix contraction, e.g.~given two $D\times D$ matrices $A, B$, the contraction $AB \approx \bar{A}\bar{B}$ where $\bar{A}$ is of dimension $D \times \chi$ and $B$ is of dimension $\chi \times D$, and the approximation is an example of a low-rank matrix factorization. The singular value decomposition (SVD) is an optimal (with respect to the Frobenius norm) low-rank matrix factorization.
Singular value decomposition is also at the heart of compressing tensor network bonds. \REVISE{For example, if we have two tensors connected by bonds (Fig.~\ref{fig:tensor-primitives}A), we can view the bonds as performing a matrix contraction (Fig.~\ref{fig:tensor-primitives}B), and use SVD to replace the connecting bonds by one of dimension $\chi$ (Fig.~\ref{fig:tensor-primitives}D). In the general tensor network setting, however, things are more complicated, because when} compressing a contraction between two tensors, one should consider the other tensors in the network, which affect the approximation error. \REVISE{The effect of the surrounding tensors on the compression of a given bond is commonly known as including the ``environment'' or ``gauge'' into the compression. We consider how to perform bond compression, including a simple way to include environment effects into the bond compression for general graphs, in section~\ref{sec:bondcompression}.}

Given a compression method, we view the approximate tensor network contraction as composed of a sequence of contraction and compression steps.
Compressions do not commute with contractions (or each other) thus a contraction tree with compression (\REVISE{an approximate} contraction tree) is an ordered tree.  An example tree $\Upsilon$ is shown in Fig.~\ref{fig:tndiagram}B, where in addition to the contraction operations (the green dots), we see compressions of bonds between tensors (the orange lines). The
 ordered sequence of contractions and compressions is visualized in Fig.~\ref{fig:tndiagram}C.
If we work in the setting where the compressed bond dimension $\chi$ is specified at the start, then
\REVISE{once the approximate contraction tree is written down,
the memory or computational cost of the contractions and compressions can be computed.
Optimizing the approximate contraction for such costs thus corresponds to optimizing over the space of approximate contraction trees.

\begin{figure}[t!]
  \centering
  \includegraphics[width=\linewidth]{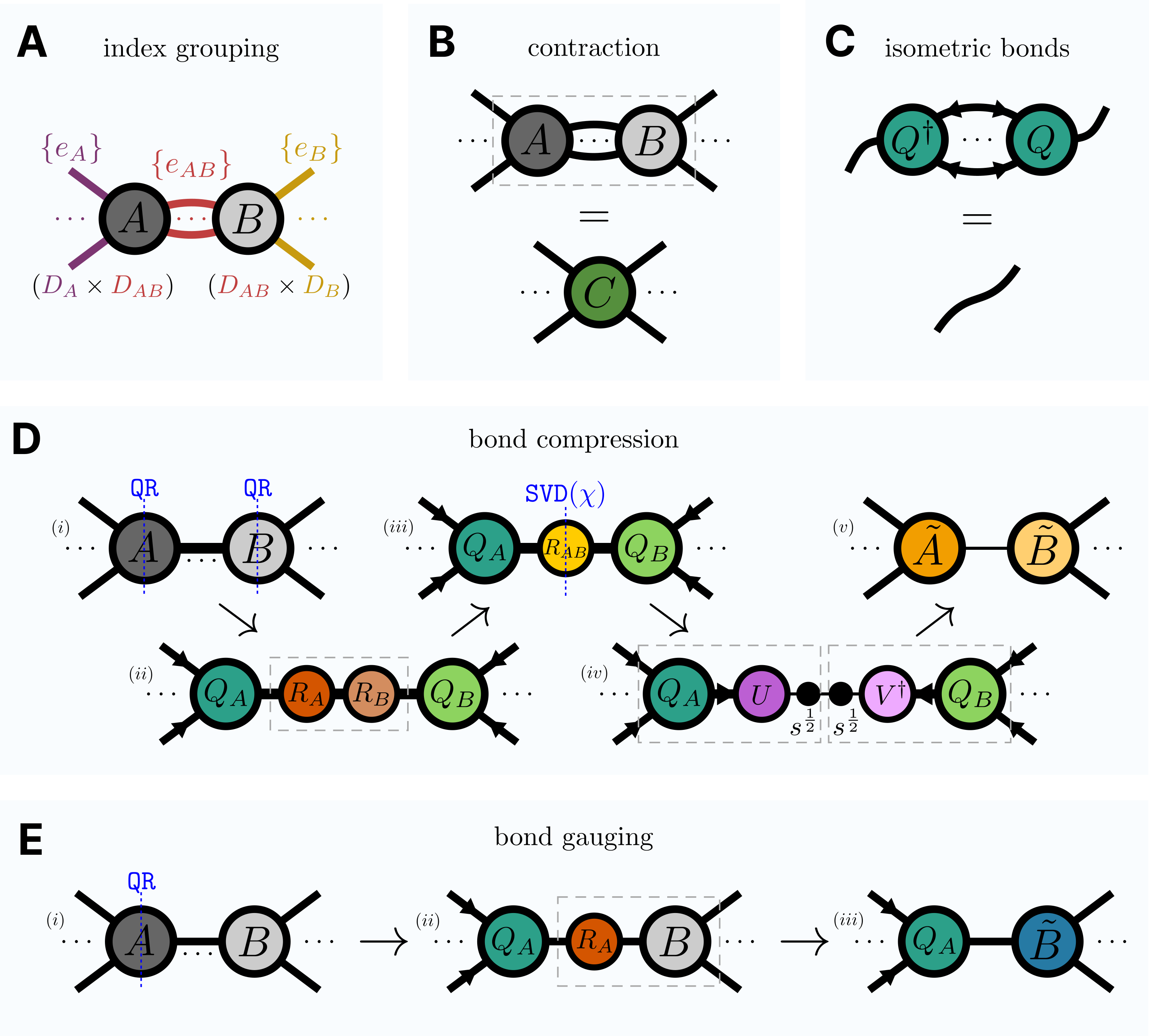}
  \caption{
    \REVISE{
  Primitive tensor operators for approximate contraction.
  \textbf{A}:
  Grouping of indices into left, shared and right sets, giving a
  matricization of the product $A B$, with rank $D_{AB}$.
  \textbf{B}
  Graphical depiction of contracting two tensors $AB \rightarrow C$.
  \textbf{C}
  Graphical depiction of an isometric tensor, $Q$, such that when dimensions
  with incoming arrows are grouped $Q^{\dagger} Q = 1$.
  \textbf{D}
  Compression of the shared bonds between two tensors $A$ and $B$, via QR
  reduction and truncated SVD to new shared bond dimension $\chi$.
  \textbf{E}
  Gauging of the bond between $A$ and $B$ to generate an isometric tensor
  $Q_A$.
  }
}\label{fig:tensor-primitives}
\end{figure}

The space of trees to optimize over is extremely large. This we tackle in two ways: by defining the position of compression steps in the tree entirely in terms of where the contractions take place (discussed in Sec.~\ref{sec:latecompression}), which means
we only need to optimize over the order of the contractions; and by using the hyper-optimization strategy, where (families) of trees are parameterized by a small set of heuristic parameters, constituting a reduced dimensionality search space (described in Secs.~\ref{sec:treegenerator},~\ref{sec:hyper}).

Ideally we wish to minimize the error of the approximate contraction as well as the cost, but the error is not known a priori. This can only be examined by benchmarking the errors of our hyper-optimized contraction trees. This is the subject of section~\ref{sec:benchmarking}.}

\REVISE{
Note that other ingredients could also be included in an
 approximate tensor network contraction algorithm, for example, the use of factorization to rewire a tensor network, generating a graph with  more vertices~\cite{levin2007tensor,pan2020contracting}; or
 inserting unitary disentanglers to reduce the compression error~\cite{evenbly2015tensor,hauru2018renormalization}. We do not currently consider these ingredients in our algorithm search space, although the framework is sufficiently flexible to include such ingredients in the future.
 We also note that some of these additional ingredients are targeted at the renormalization of loop correlations in tensor networks to yield a proper RG flow~\cite{yang2017loop}. We discuss the corner double line (CDL) model and show that it is accurately contractible with our approximate strategy in the SI~\cite{si}.
}

To aid in our discussion of the ingredients of the approximation contraction algorithm, and how to examine our choices, we will use a set of standard benchmark models, which we now discuss.

\begin{figure*}[t!]
\centering
\includegraphics[width=0.9\textwidth]{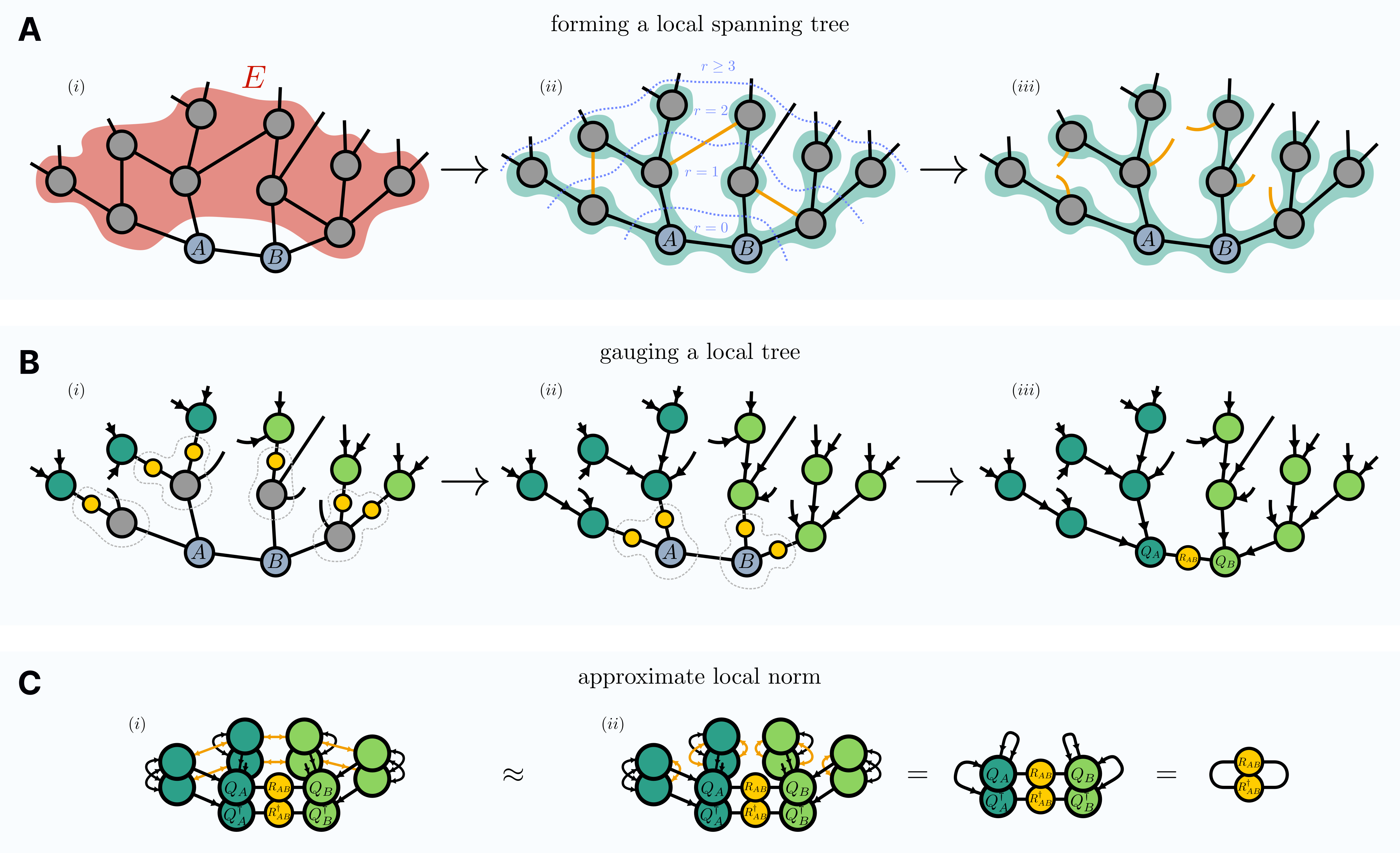}
\caption{\label{fig:tree-gauge}
\REVISE{
Overview of the tree gauge for improving bond compression accuracy, suitable for arbitrary local geometry.
\textbf{A}:
Given the bond $e_{AB}$ connecting tensors $A$ and $B$, we want to take into account
information from the surrounding environment $E$, shown in (i).
In (ii) we form a local spanning tree (shaded bonds) up to distance $r=2$ from $A$ and $B$.
If we consider `loop' bonds (colored orange) that are not part of the spanning tree
as cut, then the resulting local tree environment shown in (iii) can be optimally compressed as a proxy target.
\textbf{B}:
Depiction of the gauging process for a local tree. In (i) tensors at distance $r=2$ from $A$ and $B$ are QR decomposed, and the $R$ factors (yellow circles) are accumulated towards the bond $e_{AB}$, see Fig.~\ref{fig:tensor-primitives}E. In (ii) the same happens for $r=1$ tensors, and finally in (iii) the $R$-factors from $A$ and $B$ after accumulating all the outer gauges, $R_A$ and $R_B$, are contracted to form $R_{AB}$.
\textbf{C}:
The Frobenius norm (squared) of an $r=1$ local region in the tree gauge is shown in (i).
The norm of the local network with loop bonds (orange) cut is shown in (ii), which is exactly encoded, due to the isometric tensors, as ${({\mathrm{Tr}R^{\dagger}_{AB} R_{AB}})}^{1/2}$.
Performing a truncated SVD on $R_{AB}$ is thus only $r$-locally optimal up to the presence of such loop bonds.
}
}
\end{figure*}

\subsection{Models for testing}

To assess our algorithmic choices, we will consider two families of lattices and two tensor models. (Note that these are only the tensor networks we use for testing the algorithm; Sec.~\ref{sec:power} further considers other models to demonstrate the power of the final protocol). The two types of lattices we consider are (i) the 2D square and 3D cubic lattices, which reflect the structured lattices commonly found in many-body physics applications, and (ii) 3-random regular graphs (graphs with random connections between vertices, where each vertex has degree 3). On these lattices, the two types of tensors we consider are (i) (uniaxial) Ising model tensors, at inverse temperature $\beta$ close to the critical point, (ii) tensors with random entries drawn uniformly from the distribution $[\lambda, 1]$ (we refer to this as the {URand} model). Changing $\lambda$ allows us to tune between positive tensor network contractions and tensors with random signs, the latter case being reminiscent of some random circuit tensor networks. In all models, the dimension of the tensor indices of the initial tensor network will be denoted $D$, while the dimension of compressed bonds will be denoted $\chi$; we refer to the value of the tensor network contraction as $Z$, and the free energy per site $f = {-\ln Z}/{\beta N}$, where $N$ is the number of spins. More discussion of these models (as well as a treatment of corner double line models~\cite{yang2017loop}) is in the SI~\cite{si}.

\subsection{Bond compression strategies}\label{sec:bondcompression}

We first define how to compress \REVISE{the shared bonds $\{e_{AB}\}$} between tensors $A, B$.
We can matricize these by grouping the indices as $\{e_{A}\} / \{e_{AB}\}$, $\{e_{AB}\}$, and $\{e_{B}\} / \{e_{AB}\}$, with effective dimensions $D_A$, $D_{AB}$ and $D_B$ respectively (see Fig.~\ref{fig:tensor-primitives}A).
Generally $D_{AB} < D_A$ and $D_{AB} < D_B$ and so $AB$ is already low-rank and we can avoid forming it fully.
Instead we perform QR decompositions of the matricized $A$, $B$, giving
\begin{align}
AB &= Q_A (R_A R_B) Q_B \\
&= Q_A (R_{AB}) Q_B
\end{align}
where the $Q$ matrices satisfy the canonical conditions $Q_A^\dag Q_A=1$, $Q_B Q_B^\dag = 1$, with the canonical direction indicated by an arrow in graphical notation shown in Fig.~\ref{fig:tensor-primitives}C (detailed in the SI~\cite{si}). Then, we obtain the compressed $\bar{A}$, $\bar{B}$ through the SVD of $R_{AB}$,
\begin{align}
R_{AB} \approx U_A \sigma V_B^\dag \nonumber \\
\bar{A} = Q_A U_A \sigma^{1/2} \nonumber  \\
\bar{B} = \sigma^{1/2} V_B^\dag Q_B, \label{eq:svd}
\end{align}
truncating to $\chi$ maximal singular values in $\sigma$. Because of the canonical nature of the $Q$ matrices, truncating the SVD of $R_{AB}$ achieves an optimal compression in the matrix Frobenius norm of $AB$ due to the orthogonality of $Q_A$, $Q_B$.

\begin{figure*}[t!]
  \centering
  \includegraphics[width=1.0\textwidth]{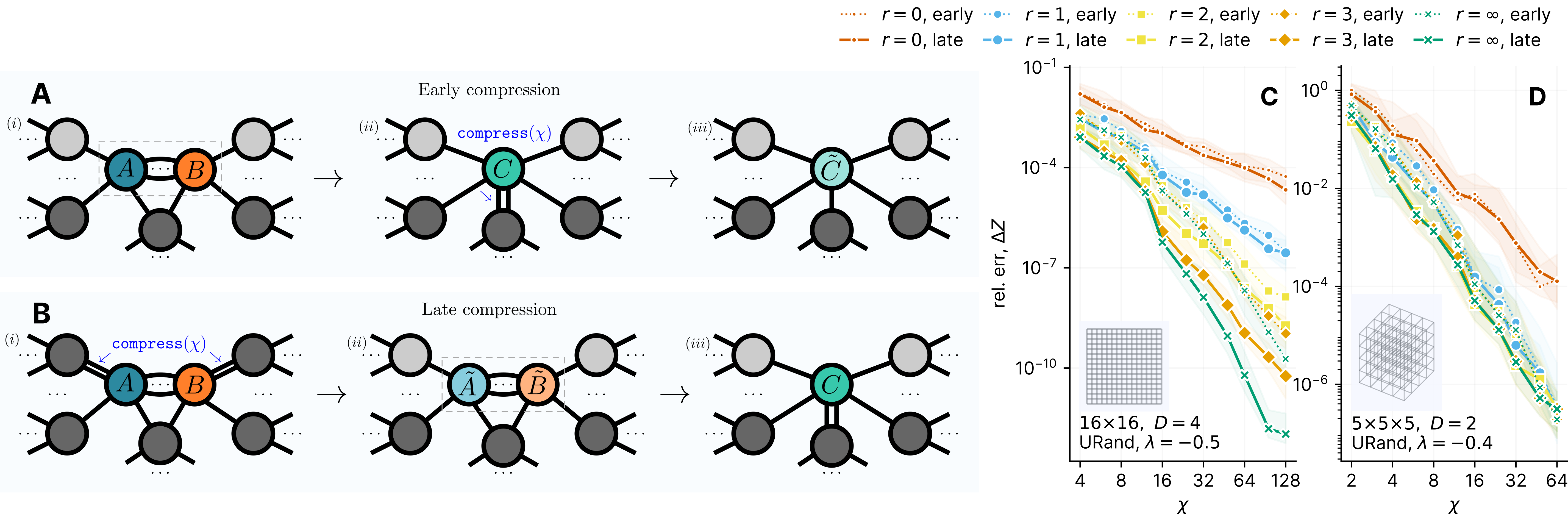}
  \caption{
  \textbf{A}: Schematic of `early' compression, where \emph{after} each pairwise contraction, any shared bonds of total size $>\chi$ are truncated.
  \textbf{B}: Schematic of `late' compression, where \emph{before} each pairwise contraction, any shared bonds of total size $>\chi$ are truncated.
  \textbf{C}: Error $\Delta Z$ of contracting a $16{\times}16$ $D=4$ TN with uniform random entries $\in [-0.5, 1]$ as a function of $\chi$, tree gauging distance, $r$, and either early or late compression.
  \REVISE{The TN is contracted using the standard MPS boundary contraction algorithm.
  Line (band) shows median (interquartile range) over 50 instances.
  \textbf{D}: The same but for an approximate contraction of a $5{\times}5{\times}5$ $D=2$ tensor network with uniform random entries
  $\in [-0.4, 1]$. The 3D TN is contracted using a hyper-optimized \texttt{Span} tree.
  }
  }\label{fig:early-vs-late}
\end{figure*}

Usually $\mathcal{T}$ will contain additional tensors connected to $A, B$. We refer to the additional network of connected tensors as the environment $E$, with $\mathcal{T} = \sum_{\{ e\}/\{ e_\text{out}\}} A_{\{ e_A\} } B_{\{e_B\}} E_{\{e_E\}}$ (Fig.~\ref{fig:tree-gauge}A). To compress the bond $e_{AB}$ optimally, we must  account for $E$. We first \REVISE{consider the case when} $E$ forms a tree around the bond $e_{AB}$ (Fig.~\ref{fig:tree-gauge}A(iii)).
Then, we can perform QR inwards from the leaves of the tree, pushing the $R$ factors towards the bond (Fig.~\ref{fig:tree-gauge}B (i)--(iii)). This is a type of gauging of the tensor network (i.e.~it changes the tensors but does not change the contraction $\mathcal{T}$) and we refer to this as setting the bond $e_{AB}$ in the tree gauge; \REVISE{alternatively, we can say the tensors in the tree are in the canonical form centered around bond $e_{AB}$.}
This results in a similar matricized $\mathcal{T} = Q_A (R_{AB}) Q_B$ where $Q_A$, $Q_B$ have accumulated the products of $R$ factors from all tensors to the left and right of bond $e_{AB}$ \REVISE{(Fig.~\ref{fig:tree-gauge}B (iii))}. Then, the truncated SVD of $R_{AB}$ in~\eqref{eq:svd} similarly achieves an optimal compression of $e_{AB}$ with respect to error in $\mathcal{T}$.

More generally, $\mathcal{T}$ may contain loops, which extend into the environment \REVISE{(Fig.~\ref{fig:tree-gauge}A)} and a similarly optimal gauge is hard to compute~\cite{evenbly2018gauge,HauruRenormalizationtensornetworks2018}. However, \REVISE{by cutting loops in the environment $E$ (i.e.~not contracting some of the bonds in the loops) we obtain} a tree of tensors around bond $e_{AB}$, e.g.~a spanning tree out to a given distance $r$. \REVISE{(There are multiple ways to cut bonds to obtain a spanning tree; the specific spanning tree construction heuristic is given in the SI~\cite{si}).}
Placing $e_{AB}$ in the tree gauge (of distance $r$), we can then perform the same compression by truncated SVD, but without the guarantee of optimality \REVISE{since we are neglecting loop correlations, see Fig.~\ref{fig:tree-gauge}C}. \REVISE{However,} this type of tree gauge compression \REVISE{is easy to use in the general graph setting, and thus} will be the main bond compression scheme explored in this work.

\REVISE{
One can show~\cite{wangClusterUpdateTensor2011,corbozCompetingStatesTJ2014,iinoBoundaryTensorRenormalization2019,si} that performing the truncation in Eq.~\eqref{eq:svd}
is equivalent to inserting the projectors $P_A{=}R_B V_B \sigma^{-1/2}$ and $P_B{=}\sigma^{-1/2} U_A^\dagger R_A$ such that $AB \approx A P_A P_B B$.
As such, having computed $R_A$ and $R_B$ from the local spanning trees we can form and contract $P_A$ and $P_B$ directly into our original tensor network without affecting any tensors other than $A$ and $B$, but still include information from distance $r$ away. In other words, the steps depicted in Fig.~\ref{fig:tree-gauge} are performed virtually, which avoids having to reset the gauge after compression.
}

\begin{figure*}[t!]
\centering
\includegraphics[width=1.00\textwidth]{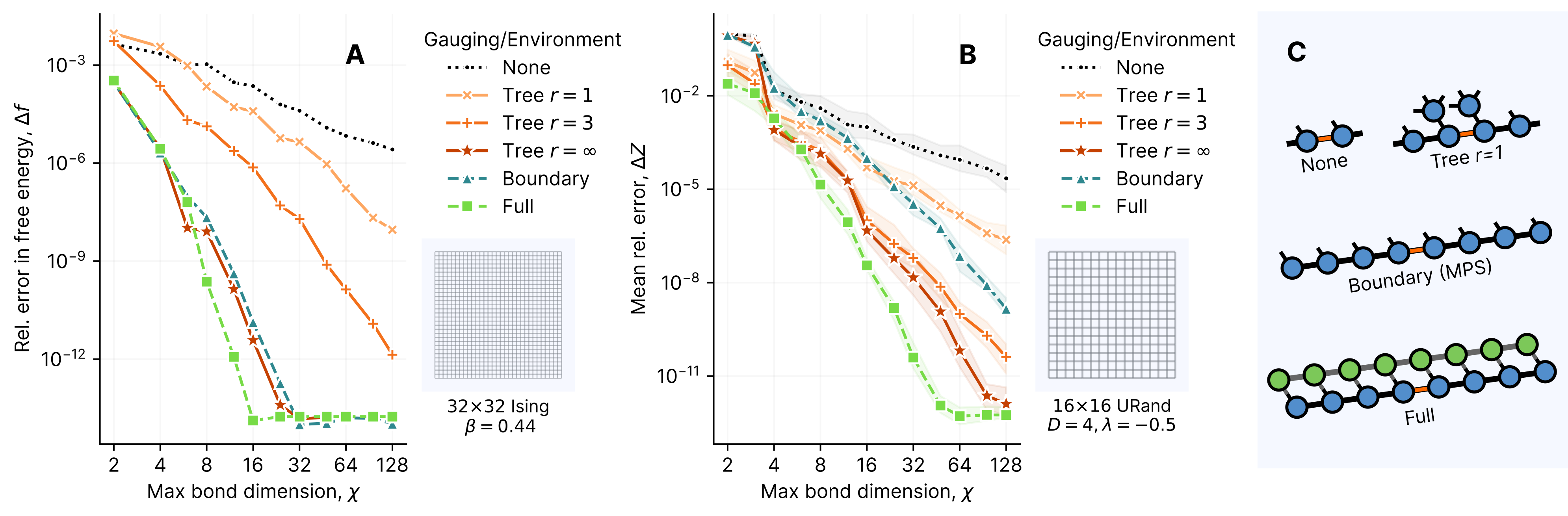}
\caption{
\textbf{A}:
Error in free energy, $\Delta f$, as a function of bond dimension, $\chi$, for different gauging and environments for the 2D Ising model at the critical point.
\textbf{B}:
Contraction error, $\Delta Z$, for the same settings but on a $D=4$ URand model with $\lambda=-0.5$. \REVISE{All contractions contract from the boundary row by row, thus all bond compressions are for bonds on the boundary. However, different gauging is performed before the compressions. None: no gauging, Tree: bonds are placed in the tree-gauge up to distance $r$, followed by `late' compression, Boundary: bonds are placed in the canonical form of the MPS boundary, before compression, Full: the environment around tensors $A$, $B$, is explicitly contracted using a counter-propagating MPS of the same bond dimension, and the bond between $AB$ is then truncated to minimize the error in $\mathrm{Tr}\ BEA$.
Note that since $E$ is itself only approximate and many truncations are compounded the error overall is not guaranteed to be smaller than another method -- as seen for some small $\chi$ points here.
}
\textbf{C}: Illustration of the different environments that a single compression step is optimal with regard to.
}\label{fig:fig-error-vs-gauging-2d}
\end{figure*}

\subsection{Early versus late compression}\label{sec:latecompression}

In practice, compression must be performed many times during a tensor network contraction. It might seem natural to perform compression immediately after two tensors are contracted to form a tensor \REVISE{larger than some size threshold, here given by a maximum bond dimension $\chi$} (early compression). This is illustrated in Fig.~\ref{fig:early-vs-late}A. However, as discussed above, including information from the environment is important for the quality of compression. Early compression means that tensors in the environment are already compressed, decreasing their quality. An alternative strategy is to compress a bond between tensors only when one of them \REVISE{(exceeding the size threshold)} is to be contracted (late compression), as illustrated in Fig.~\ref{fig:early-vs-late}B. By delaying the compression, more bonds/tensors in the environment are left uncompressed, \REVISE{which can potentially improve the quality of the contraction. However, late compression will also increase the cost/memory of contraction (as there are more large tensors to consider). This means that it is most efficient to use late compression when the associated gain in accuracy is large.}

In Fig.~\ref{fig:early-vs-late}C, we assess the effect of early versus late compression when contracting a 2D $16\times 16$ lattice ($D=4$, URand model with \REVISE{tensor} entries $\in [-0.5, 1]$). All compressions are performed using the tree gauge (out to some distance $r$, several tree distances $r$ are shown), \REVISE{and we show the relative error of the contraction $\Delta Z$ as a function of the maximum allowed bond dimension $\chi$.}  We see in this case that late compression is more accurate than early compression, and that this improvement  increases when using larger tree gauge distances, reflecting the fact that the gauging is incorporating more environment information. \REVISE{In Fig.~\ref{fig:early-vs-late}D, we similarly compare early versus late compression using the tree gauge in a 3D lattice (using a hyper-optimized \texttt{Span} tree as described later). In contrast to the 2D result, here we see a smaller improvement from performing late versus early compression and from increasing the tree gauge distance. This suggests that incorporating the effect of the environment requires a more sophisticated gauging strategy in 3D. In general, we summarize our findings as so: late compression is preferred when trying to maximize accuracy for a given bond dimension $\chi$ or size of the largest single tensor operation, while early compression can be better when optimizing computational total cost or memory for a given accuracy. In our subsequent calculations, we will indicate the choice of early or late compression in the simulations.}

\subsection{Comparison of the tree gauge to other gauges}\label{sec:assess-tree}

\REVISE{To evaluate the quality of the tree gauge compared to other gauging/environment treatments in the literature, we consider contractions on a 2D lattice.
To isolate the comparison to only the choice of gauge, we use the same approximate contraction tree as used in boundary contraction, namely contraction occurs row by row starting from the bottom, and compression occurs left to right after the entire row is contracted. We then use 4 different gauges/environment treatments during the compression: None, Tree, Boundary, and Full. None corresponds to no gauging. Tree is the tree gauge discussed above (up to distance $r$).
Boundary corresponds to the standard MPS boundary gauging~\cite{verstraete2008matrix,ranLectureNotesTensor2020}, where, after the new row of tensors has been contracted into the boundary, the boundary MPS is canonicalized around the leftmost tensor and then compressed left to right in an MPS compression sweep (see Fig. 3 of the SI for an illustration). Full corresponds to explicitly computing the environment $E$ by approximate contraction (using the standard MPS boundary contraction algorithm to contract rows from the top). Then, for the tensors $A$, $B$ sharing bond $e_{AB}$ to be compressed, the scalar value of the tensor network is $Z = \mathrm{Tr}\ B E A$ (where $A, B, E$ have been matricized). Using the eigenvector decomposition, $BEA = L \sigma R^\dag$, where $L$, $R$ are left, right eigenvectors respectively, then $e_{AB}$ is optimally compressed by defining $\tilde{B} = \tilde{L}^\dag B$, $\tilde{A} = A\tilde{R}$, where $~\tilde{L}$, $\tilde{R}$ are the eigenvectors corresponding to the eigenvalues of largest absolute magnitude~\cite{xieSecondRenormalizationTensorNetwork2009,zhaoTensorNetworkAlgorithm2015}. Note that the full environment gauge is
 expensive, as it requires an estimate of $E$ from all the tensors in the network.}

The numerical performance of the different strategies is shown in
Fig.~\ref{fig:fig-error-vs-gauging-2d} for two problems: a $32\times 32$ lattice (2D Ising model, near critical) and a $16\times 16$ lattice ($D=4$, URand model with entries $\in [-0.5, 1]$).
In all cases, we see that including some environment information is better than not including any environment (``None''). In the 2D Ising model, as the tree distance $r$ increases, tree gauge compression converges in quality to the \REVISE{MPS} boundary environment scheme (``Boundary''); \REVISE{the two are related as the MPS boundary corresponds to setting an infinite tree distance $r$ for a tree that grows only along the boundary.} In the 2D URand model, even for small $r$, the tree gauge already improves on the boundary environment. The full environment treatment yields the best compression quality for larger $\chi$, but this is achieved at larger cost.

Our numerical results in 2D suggest that the tree gauge is a reasonable compromise between accuracy and efficiency, equaling or outperforming the common boundary environment strategy, while being well-defined for more general graphs.

\subsection{Approximate contraction algorithm}

\begin{figure}[t!]
\begin{algorithm}[H]
\caption{Approximate contraction}\label{alg:compressed-contraction}
\begin{algorithmic}
\State\textbf{Input:} tensor network $\mathcal{T}$, ordered contraction tree $\Upsilon$, maximum bond dimension $\chi$, tree gauge distance $r$, flag \texttt{compress\_late}
\State// $i, j, k, l$ label tensors, $T^{[i]}, \ldots$ in $\mathcal{T}$.
\For{$i, j \in \Upsilon$}
\If{$\texttt{compress\_late}$}
\For{$l \in \textsc{neighbors}(\mathcal{T}, i)$}
\If{$\textsc{bondsize}(\mathcal{T}, i, l) > \chi ~\textbf{and}~ l \neq j$ }
\State$\textsc{compress\_tree\_gauge}(\chi, r, \mathcal{T}, i, l)$
\EndIf
\EndFor
\For{$l \in \textsc{neighbors}(\mathcal{T}, j)$}
\If{$\textsc{bondsize}(\mathcal{T}, j, l) > \chi ~\textbf{and}~ l \neq i$ }
\State$\textsc{compress\_tree\_gauge}(\chi, r, \mathcal{T}, j, l)$
\EndIf
\EndFor
\EndIf
\State$k \leftarrow \textsc{contract}(\mathcal{T}, i, j)$
\If{\textbf{not}~$\texttt{compress\_late}$}
\For{$l \in \textsc{neighbors}(\mathcal{T}, k)$}
\If{$\textsc{bondsize}(\mathcal{T}, k, l) > \chi$ }
\State$\textsc{compress\_tree\_gauge}(\chi, r, \mathcal{T}, k, l)$
\EndIf
\EndFor
\EndIf
\EndFor
\State\textbf{Return:} $\mathcal{T}$
\end{algorithmic}
\end{algorithm}
\end{figure}

Given a choice of late or early compression, and using the tree gauge, we can explicitly write down a simple pseudo-code version of the core approximate contraction function, Algorithm~\ref{alg:compressed-contraction}, which implements Fig.~\ref{fig:hyper-compressed-overview}B.
The exact form of the inner functions is detailed in the SI~\cite{si}. An alternative, that might be useful in some contexts, is to use the compression locations to transform a tensor network into an approximately equivalent but exactly contractible form, by inserting a set of explicit projectors -- this is also detailed in the SI~\cite{si}.

\subsection{Generating contraction trees}\label{sec:treegenerator}

After fixing the choice of early or late compression, the subsequent location of compressions in the contraction tree is purely determined by the contraction order. \REVISE{This is a major simplification, because}, when optimizing over the approximate contraction trees we need only optimize the order of contractions. Nonetheless, the space of ordered trees is still extremely large and hard to sample fully.

\REVISE{To simplify the search, we work within a lower-dimensional parameterization of the search space by introducing tree generators. These heuristics} generate trees within three structural families  we term \texttt{Greedy}, \texttt{Span}, and \texttt{Agglom}. The specific instance of tree within each family is defined by a set of hyperparameters that can then be optimized.
Here we describe the heuristic generators \REVISE{at a high level} (with a more detailed description in the SI~\cite{si}).
The input to the generators is only the tensor network graph, bond sizes and $\chi$ -- the tensor entries are not considered.

\begin{figure}[tb]
  \centering
  \includegraphics[width=\linewidth]{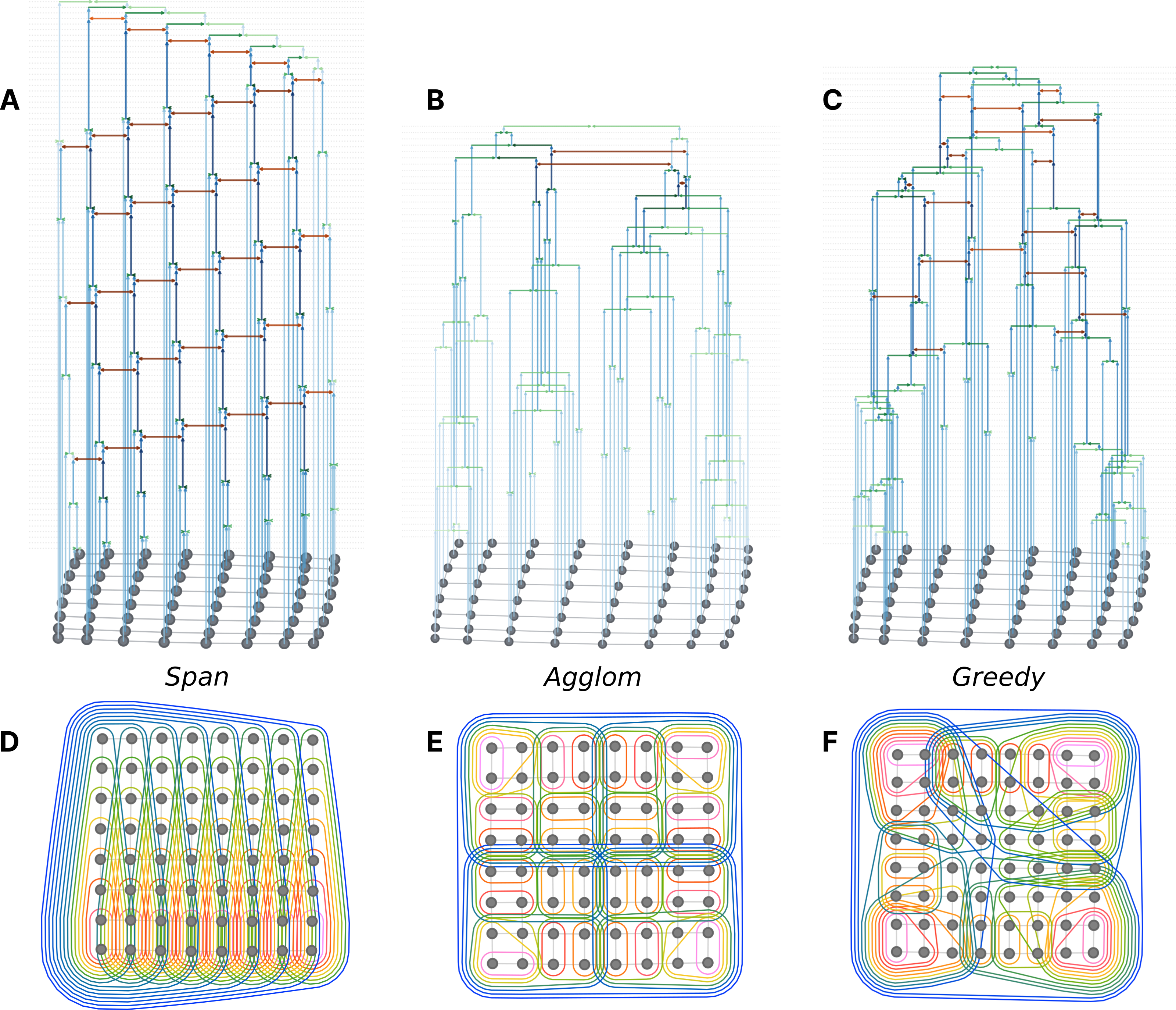}
  \caption{
  Example approximate contraction trees for a 2D $8{\times}8$ square lattice TN. {\textbf{A}}, {\textbf{B}}, and {\textbf{C}} show the ordered contraction trees with compressions (orange, using the `late' strategy) explicitly marked for the \texttt{Span}, \texttt{Agglom} and \texttt{Greedy} methods respectively. \REVISE{The computation proceeds from the bottom to top.}
  {\textbf{D}}, {\textbf{E}}, and {\textbf{F}} show the same contraction trees as hierarchical communities, with the contraction ordering proceeding from the smallest pink bands to largest blue bands.
  }\label{fig:example-trees}
\end{figure}

The \texttt{Greedy} tree generator assigns a score to each bond in the 
$\mathcal{T}$. It then chooses the highest scoring bond, generates a new $\mathcal{T}$ by simulating bond contraction and compression (i.e.\ computing the new sizes and network structure) and repeats the process, building an ordered tree. The bond score is a combination of the tensor sizes before and after (simulated) compression and contraction, a measure of the centrality of the tensors (their average distance to every other tensor), and the subgraph size of each intermediate (i.e. \REVISE{how many tensors were contracted to make the current tensor}); the hyperparameters are the \REVISE{linear} weights of each component in the score.

\begin{figure*}[t!]
  \centering
  \includegraphics[width=0.8\textwidth]{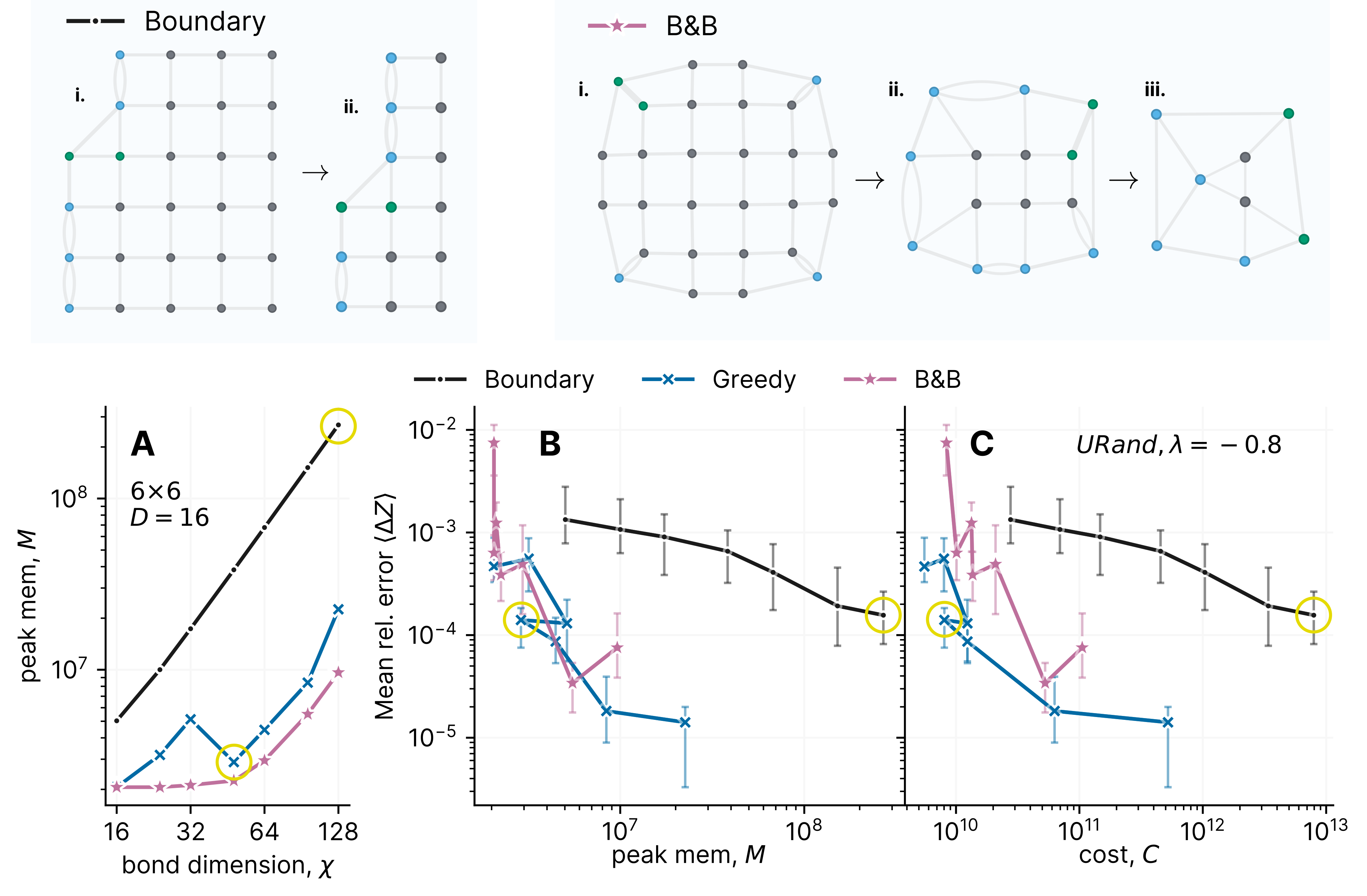}
  \caption{
  Performance of hyper-optimized approximate contraction for a  $6{\times}6$ square TN with all bonds of size $D=16$. \REVISE{The TN is filled with uniformly random entries $\in[\lambda=-0.8, 1.0]$ (2D URand model). We test 3 different contraction trees: standard MPS boundary contraction, optimization over \texttt{Greedy} trees, brute force optimization over all approximate contraction trees (B\&B).
  The upper insets show snapshots illustrating the `Boundary' contraction, an example of \texttt{Greedy}, and the optimal `B\&B' contraction for $\chi=32$.
  \textbf{A}: The peak memory usage, $M$, of the \REVISE{standard MPS `boundary'} method, compared to the \texttt{Greedy} and `B\&B' algorithms which have been optimized for each $\chi$.
  \textbf{B} and \textbf{C}: the error for each plotted against peak memory, $M$, and computational cost, $C$, respectively, averaged over 20 instances of the URand model. The compressions are performed late using a tree-gauge distance $r=1$. The yellow circles correspond to approximately equal error $\Delta Z \approx 10^{-4}$, using the `Boundary' and \texttt{Greedy} trees, to enable the ratio of costs to be determined; e.g.
  the observed speed-up of \texttt{Greedy} over `Boundary' for this error is $120\times$.
  }
  }\label{fig:small-example}
\end{figure*}

The \texttt{Span} tree generator is inspired by boundary contraction. It generates a directed spanning tree of the original graph, and the contraction order is chosen to contract the leaves inwards. \REVISE{Contracting simultaneously along all the branches of the tree
defines an effective contraction boundary, that sweeps towards the root.}
The algorithm begins by choosing either the most or least central node in the TN, $i_0$, as an initial span, $\mathcal{S} = \{i_0\}$. It then greedily expands to a connected node $j$, adding the contraction $(i, j) \rightarrow i$ in \emph{reverse} order to the path, where $i \in \mathcal{S}, j\notin \mathcal{S}$.
The algorithm repeats by then considering all neighbors of the newly expanded span $\mathcal{S} \rightarrow \mathcal{S} \cup \{j\}$.
A few local quantities -- connectivity to $\mathcal{S}$, dimensionality, centrality, and distance to $i_0$ -- are combined into a score used in the greedy selection of the next node in the tree, and the combination weights are the hyperparameters.

The previous approaches grow ordered trees  locally. The \texttt{Agglom} tree generator explicitly considers the  full TN from the start and is inspired by renormalization group contraction strategies in the literature~\cite{levin2007tensor}. Given a community size $K$, the generator performs a balanced partitioning over the $N$ tensors in $\mathcal{T}$ to find $N / K$ roughly equal subgraphs.
These subgraphs then define intermediate tensors, and the tensors within the subgraph are contracted using the \texttt{Greedy} algorithm with default parameters. After simulating the sequence of compressions and contractions, the network of intermediate tensors defines a new ``coarse grained'' tensor network for which the agglomerative process can be repeated. In this work, \texttt{Agglom} uses the KaHyPar graph partitioner~\cite{kahypar1,kahypar2}, treating the community size $K$, imbalance, partitioning mode and objective as the tunable hyperparameters.

Some sample ordered contraction trees generated by the above heuristics are shown in Fig.~\ref{fig:example-trees} for a 2D $8 \times 8$ lattice. In particular, we observe the boundary-like contraction order of the \texttt{Span} tree (contracting row by row \REVISE{from the bottom}) and the hierarchical RG like structure of the \texttt{Agglom} tree (forming increasing clusters); the \texttt{Greedy} tree contracts simultaneously from all 4 corners inwards, rather than from one side like the \texttt{Span} tree. Note that the \texttt{Agglom} tree tends to perform  more contractions before compressions are performed than the \texttt{Span} tree because it constructs many separate clusters simultaneously, and the \texttt{Greedy} tree exhibits behavior intermediate between the two.

\subsection{Optimizing the contraction trees}\label{sec:hyper}

We optimize the trees by tuning the hyperparameters that generate them with respect to a cost function.
Since we also wish to sample many different trees it is important that the cost function is cheap to evaluate. We perform the optimization over the hyperparameter space using Bayesian optimization~\cite{optuna_2019,nevergrad}, which is designed for gradient free high dimensional optimization.
The overall process is shown in Fig.~\ref{fig:hyper-compressed-overview}A, with more detailed pseudo-code in the SI~\cite{si}.

Depending on the computational resources available, we can choose the cost function to be memory (peak memory usage $M$) or the computational (floating point) cost $C$. For $C$ we include the cost of contractions, QR and SVD decompositions.

We optimize the contraction trees over the hyperparameters in each of the 3 families of ordered tree generators. In all results with optimized trees, we used a budget of 4096 trees, though in practice a few hundred often achieves the same result.
\REVISE{
The practical effect of the hyper-optimization time is considered in the SI.
}

\begin{figure*}[t!]
  \centering
  \includegraphics[width=0.8\textwidth]{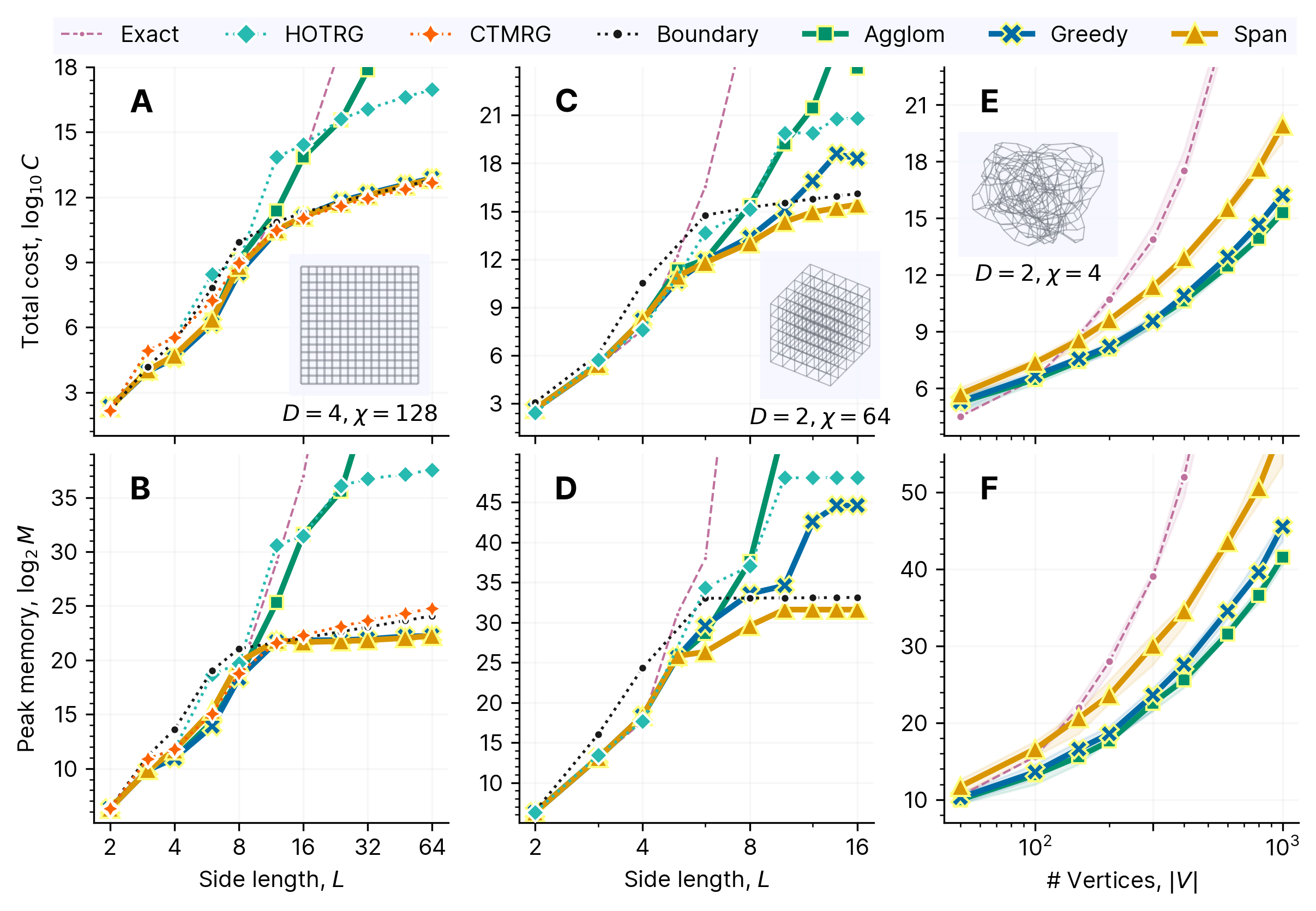}
  \caption{
  Contraction peak memory and cost of approximate contraction trees in different families, versus exact contraction cost and \REVISE{CTMRG, HOTRG and boundary  contraction algorithms, for different geometries (insets show sample geometry).}
  The tree gauging distance is set to $r=0$ for this comparison, with `early' compression, and we also turn off gauging in the Boundary algorithm.
  \REVISE{
  The hyper-optimized trees are optimized for $M$ and $C$  separately.
  }
  \textbf{A} and \textbf{B}: peak memory $M$, and cost $C$, of contracting a 2D square TN with $D=4$ and $\chi=32$ as a function of side length $L$.
  \textbf{C} and \textbf{D}: peak memory $M$, and cost $C$, of contracting a 3D cube TN with $D=4$ and $\chi=32$ as a function of side length $L$.
  \textbf{D} and \textbf{E}: peak memory $M$, and cost $C$, of contracting 3-regular random graphs with initial bond dimension $D=2$ and $\chi=4$ as a function of number of vertices $|V|$.
  The line and shaded band show the median and interquartile range across 20 instances respectively.
  }\label{fig:tree-vs-geom-complexity}
\end{figure*}

\begin{figure}[t!]
    \centering
    \includegraphics[width=0.9\linewidth]{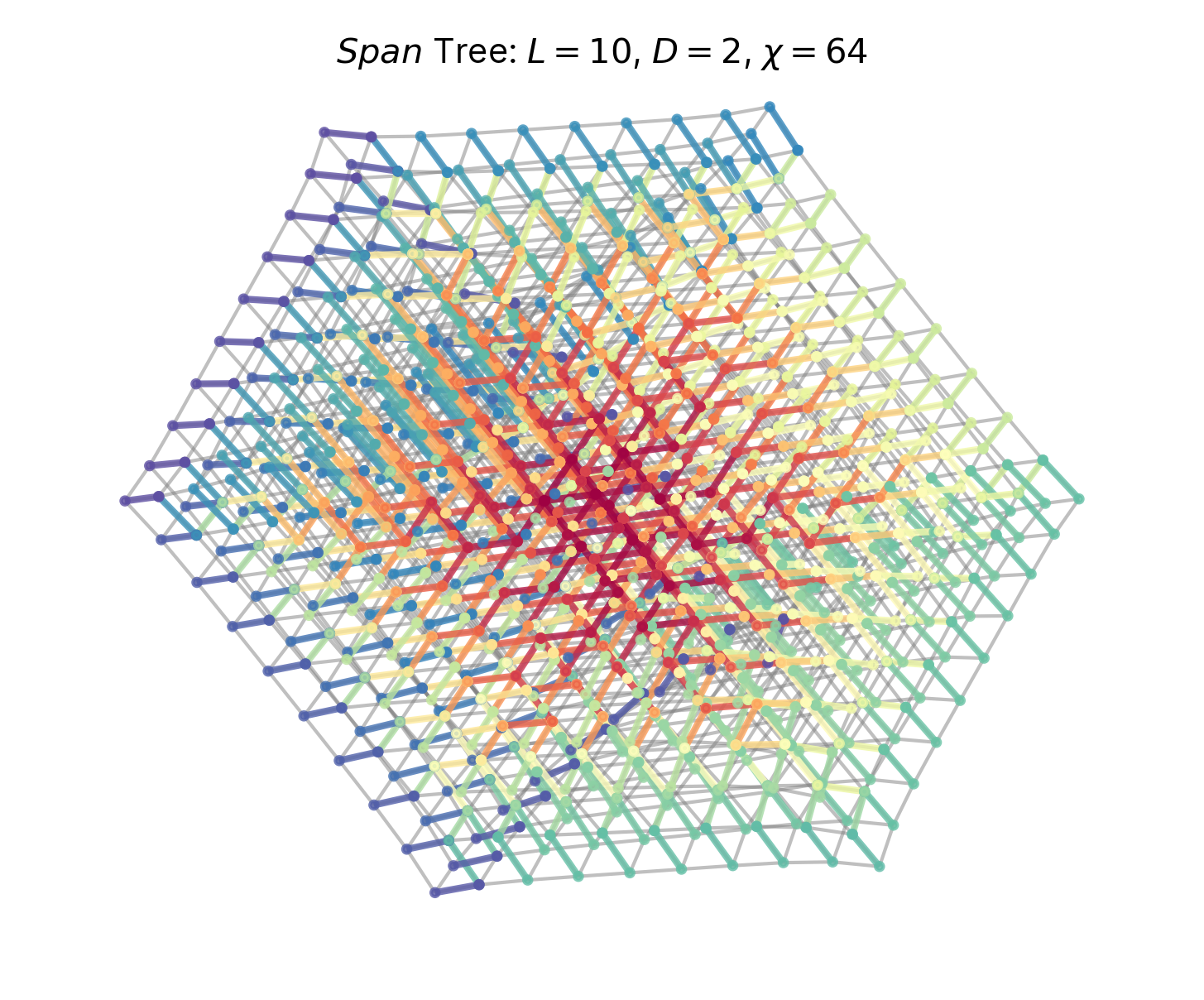}
    \caption{
    \REVISE{
    Depiction of the contraction tree found by the \texttt{Span} algorithm, optimized for minimum floating point cost $C$, for a large 3D lattice. The colors of the highlighted edges and nodes indicate the stage of contraction they are involved in, running from blue (earliest) to red (latest).
    }
    }\label{fig:span3dcontraction}
\end{figure}

\begin{figure*}[t!]
  \centering
  \includegraphics[width=0.8\textwidth]{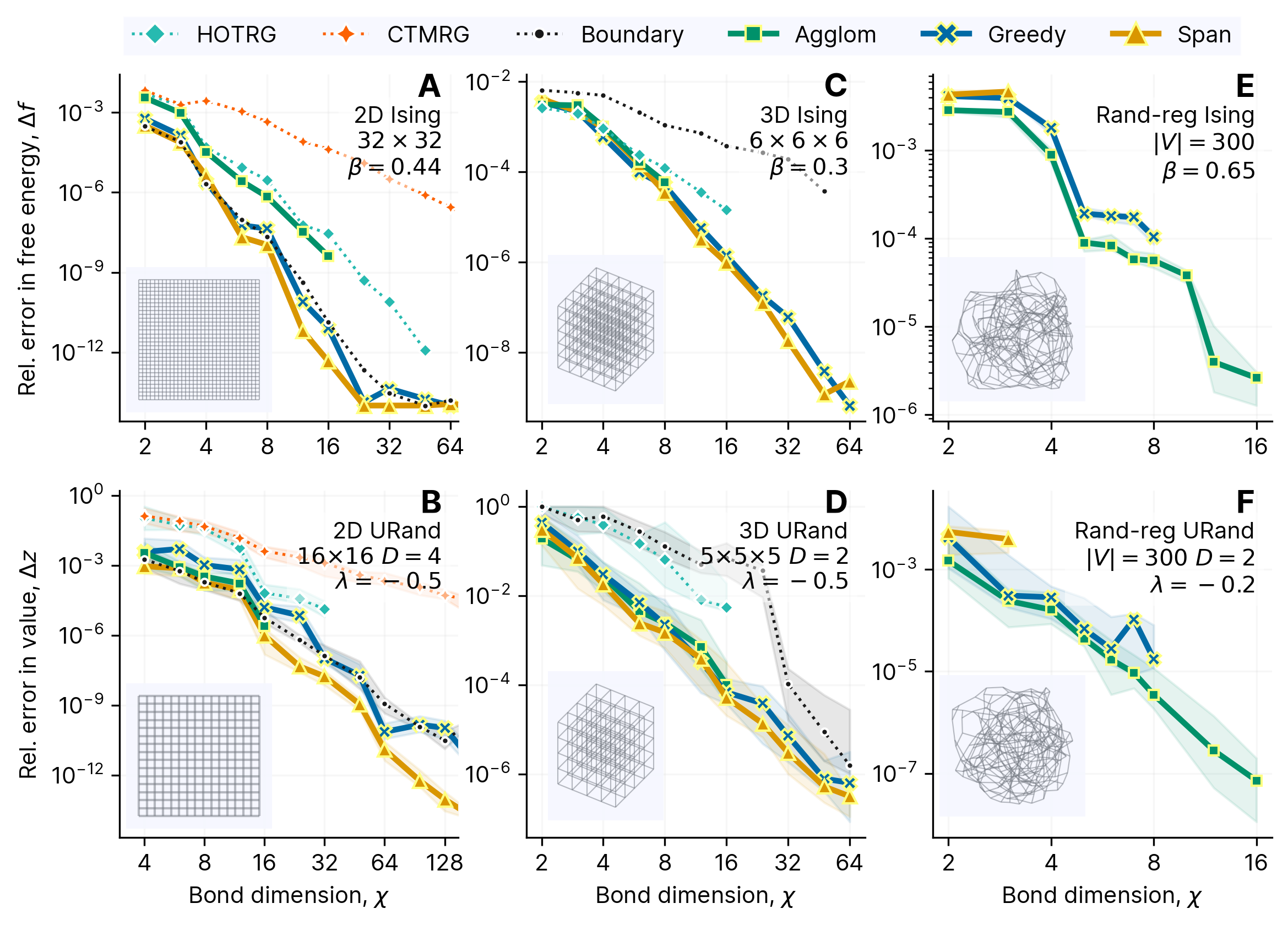}
  \caption{
Error versus compressed bond dimension $\chi$ of hyper-optimized approximate contraction using optimized \texttt{Span}, \texttt{Greedy}, and \texttt{Agglom} trees, in comparison to boundary contraction, CTMRG, and HOTRG, for medium size TNs.
(Insets show sample geometry).
  In terms of gauging settings for the hyper-optimized methods, in 2D we use $r=6$ with late compressions, and for the others we use $r=3$ and early compression.
  \textbf{A}: Relative error in the free energy of the 2D Ising model on a $32\times32$ square lattice close to the critical point.
  \textbf{B}: Relative error in the contracted value of the 2D URand model on a $16\times16$ square lattice with $D=4$ in the intermediate hardness regime of $\lambda$.
  \textbf{C}: Relative error in the free energy of the 3D Ising model on a $6 \times 6 \times 6$ cubic lattice close to the critical point.
  \textbf{D}: Relative error in the contracted value of the 3D URand model on a $5\times5\times5$ cubic lattice with $D=2$ in the intermediate hardness regime of $\lambda$.
  \textbf{E}: Relative error in the free energy of the Ising model on 3-regular random graphs with $|V|=300$ close to the critical point (line and bands show median and interquartile range across 20 instances).
  \textbf{F}: Relative error in the contracted value of the URand model on 3-regular random graphs with $D=2$ in the intermediate hardness regime of $\lambda$ (line and bands show median and interquartile range across 20 instances).
  }\label{fig:err-vs-chi}
\end{figure*}

\begin{figure*}[t!]
    \centering
    \includegraphics[width=\textwidth]{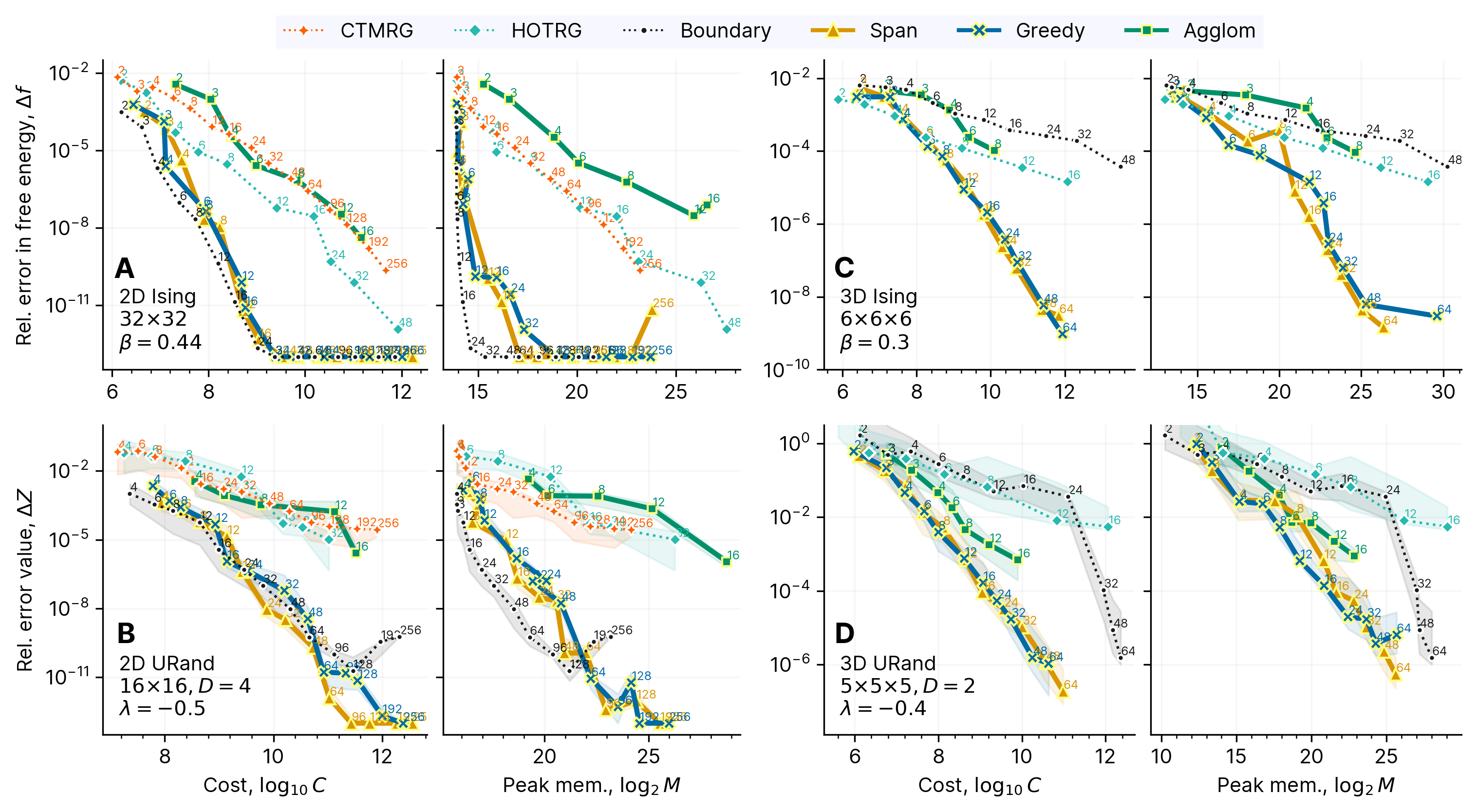}
    \caption{
    \REVISE{
        Error versus cost of hyper-optimized approximate contraction using optimized \texttt{Span}, \texttt{Greedy}, and \texttt{Agglom} trees, in comparison to boundary contraction, CTMRG, and HOTRG, for medium size TNs where the exact reference values are available.
        Here the error is plotted against either the total cost of the contraction, $C$, or peak memory requirement $M$, as computed by tracing through the computation.
        The trees are optimized for each separately.
        Four different lattice and model combinations are shown:
        \textbf{A} -- 2D Ising model,
        \textbf{B} -- 2D URrand model,
        \textbf{C} -- 3D Ising model, and
        \textbf{D} -- 3D URrand model.
        The lines are annotated with the value of $\chi$.
        The gauging settings used for the hyper-optimized methods are $r=6$ and late compression in 2D and $r=2$ and early compression in 3D.
        }
    }\label{fig:performance-ctmrg-hotrg-small}
\end{figure*}

\subsection{Quality of hyper-optimization}

To test the quality of the hyper-optimization and the tree search space, we first consider a small (but nonetheless non-trivial-to-contract due to large $D$ and $\chi$) square TN of size $6 \times 6$. In this case, the number of tensors is small enough that it is possible to perform an exhaustive search over all ordered contraction trees using a branch and bound algorithm (`B\&B' -- see SI);
\REVISE{
here we  minimize peak memory $M$.
}
In Fig.~\ref{fig:small-example} we compare the performance of the hyper-optimized \texttt{Greedy} algorithm against the exhaustive branch and bound search \REVISE{(using tree gauging for compression in each case for a fair comparison). The standard boundary contraction algorithm is also shown as a comparison point.}

As shown in Fig.~\ref{fig:small-example}A, \REVISE{hyper-optimizing over the space of \texttt{Greedy} trees produces performance quite similar to the B\&B search and very different from the standard boundary contraction. This indicates that the hyper-optimization is doing a good job of searching the approximation contraction tree space for graphs of this size. In the top-panel, we can see the optimal contraction strategy found by B\&B produces a very different contraction order to boundary contraction, exploiting the finite size of the graph and the targeted $\chi$ to significantly reduce $M$.}

We can also verify that optimizing $M$ leads to reduced  error. In Figs.~\ref{fig:small-example}B and C, we show the contraction error for the URand model with $\lambda=-0.8$, where it can be seen that for equivalent \REVISE{error ($\sim$~$10^{-4}$, indicated by the yellow circles) the peak memory or cost of using the hyper-optimized \texttt{Greedy} or B\&B approximate contraction trees is indeed much lower than that of boundary contraction.} Interestingly, the heavily optimized `B\&B' tree does not improve on the error of the \texttt{Greedy} tree for a given peak memory $M$.

\section{Benchmarking hyper-optimized approximate contraction trees}\label{sec:benchmarking}

\subsection{Summary of hand-coded strategies for regular lattices}\label{sec:comparison}

\REVISE{In our benchmarking below, when considering regular lattices, we will compare
to a range of  hand-coded contraction strategies used in the literature in many-body physics applications, namely boundary contraction, corner transfer renormalization group (CTMRG)~\cite{nishino1996corner}, and higher-order TRG (HOTRG)~\cite{xie2012coarse}. We briefly summarize the handcoded strategies here. Boundary contraction (as already used above) is a standard method in 2D, but has not been widely applied in 3D. We define a 3D version of (PEPS) boundary contraction on a cube that first contracts from one face of the cube towards the other side, leaving a final 2D PEPS tensor network that is contracted by 2D boundary contraction (with the same $\chi$). CTMRG is usually applied in 2D and to infinite systems. Here we apply CTMRG to the finite lattice by using a finite number of CTMRG moves~\cite{si}. Finally, HOTRG has been applied to both 2D and 3D infinite simulations; here we perform a limited number of RG steps appropriate for the finite lattice.
For both CTMRG and HOTRG, we also compute and insert different projectors for each compression, since we are dealing with generically in-homogeneous systems.
Illustrations of all the algorithms are given in the SI~\cite{si}.}

\begin{figure*}[t!]
    \centering
    \includegraphics[width=\textwidth]{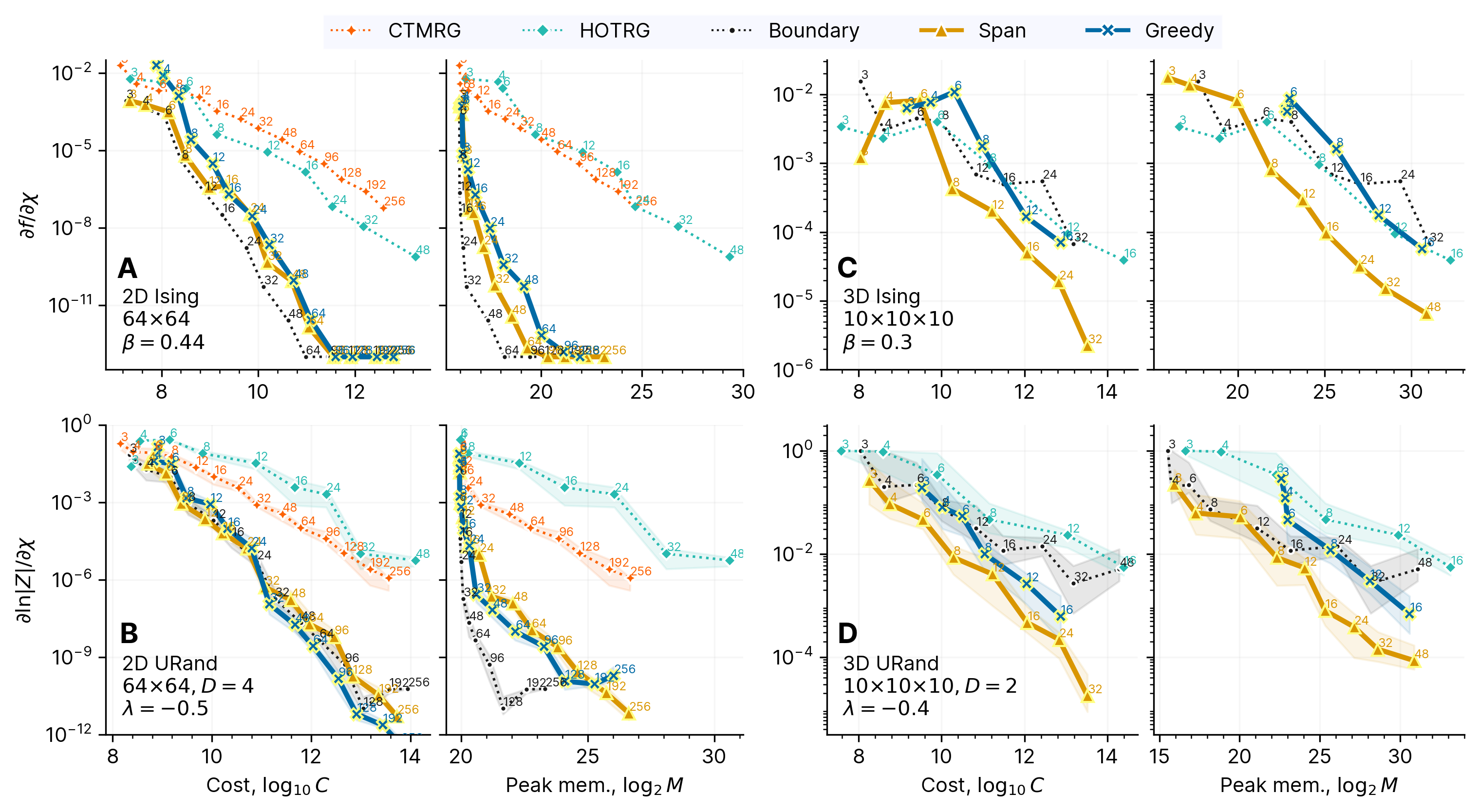}
    \caption{
    \REVISE{
        Error versus cost of hyper-optimized approximate contraction using optimized \texttt{Span}, \texttt{Greedy}, and \texttt{Agglom} trees, in comparison to  boundary contraction, CTMRG, and HOTRG, for large TNs where no exact reference is available.
        As a proxy for error, we monitor the rate of change of the log contraction value or free energy with bond dimension, $d \ln |Z|/d\chi$, $df/d\chi$, respectively.
        We plot this against both contraction cost and peak memory, where the trees have been optimized separately for each.
        The hyper-optimized methods use gauging settings of $r=6$ in 2D and $r=2$ in 3D, both with early compression.
        }
    }\label{fig:performance-ctmrg-hotrg-large}
\end{figure*}

\subsection{Cost scaling with graph size}

In Fig.~\ref{fig:tree-vs-geom-complexity} we show the computational cost (memory and floating point cost) of hyper-optimized trees in the \texttt{Greedy}, \texttt{Span}, and \texttt{Agglom} classes for a 2D square of size $L{\times}L$, a 3D cube of size $L{\times}L {\times}L$, and for 3-regular random graphs with $|V|$ vertices, using the early compression strategy, and given bond dimension $\chi$.  \REVISE{We compare against the cost of contraction trees generated by boundary contraction, CTMRG, and HOTRG.}

\REVISE{From this and other examples,
 we can make some general observations. First,  \texttt{Span} trees yield good costs for simple lattices which have a regular local structure, the \texttt{Agglom} tree is superior for random graphs, and the \texttt{Greedy} tree works well for both sets. The markedly different performance of the \texttt{Agglom} and \texttt{Span} trees on random versus simple lattices suggests that these are two good limiting cases for testing contraction heuristics.
Interestingly, in the 3D cubic case, the hyper-optimized \texttt{Span} tree performs a boundary-like contraction, but rather than contracting from one face across to the other side, it can find a strategy that contracts all faces towards a point, as visualized in Fig.~\ref{fig:span3dcontraction}. This substantially improves over the hand-coded boundary PEPS strategy in terms of cost.
Similar observations apply to the \texttt{Greedy} tree, which is similar to or outperforms both \texttt{Span} and handcoded algorithms for smaller structured lattices, although its performance degrades for larger lattices. We also find that \texttt{Greedy} trees optimize the cost function less well in other instances of large lattices
. This suggests that the search space generated by the \texttt{Greedy} tree generator is limiting at larger lattice sizes.
}
\REVISE{In the 2D square lattice, we find that compared to the hand-coded algorithms, the \texttt{Span} and \texttt{Greedy} trees with early compression are superior with respect to memory and cost, even beating out the most widely used boundary MPS strategy. At smaller system sizes,
the superior performance of \texttt{Span} and \texttt{Greedy} over boundary MPS reflects the ability of these algorithms to exploit edge and boundary effects. Interestingly, CTMRG is also superior to boundary MPS at small system sizes, and the optimized strategies seem to interpolate between a more CTMRG-like and MPS boundary-like contraction. The HOTRG algorithm exhibits similar performance to the \texttt{Agglom} tree, as expected due to its real-space RG motivated ordering of contractions. At larger sizes, boundary, CTMRG, \texttt{Span}, and \texttt{Greedy} show similar asymptotic cost, with the optimized strategies retaining a modest asymptotic improvement in memory.}

\subsection{Error versus bond dimension}\label{sec:err_vs_bond}

\REVISE{As discussed above, we optimize over the generated contaction trees for a given bond dimension $\chi$. In Fig.~\ref{fig:err-vs-chi}, we plot the
 relative error in the contraction value $\Delta Z$/free energy per site $\Delta f$ for 2D and 3D Ising and random tensor models for the hyper-optimized contraction and handcoded strategies  as a function of bond dimension.

It is natural to expect the error of an approximate contraction to decrease as we increase $\chi$, since in the limit $\chi{\rightarrow}\infty$ the algorithm becomes exact. For all the models and algorithms investigated we find a roughly polynomial suppression of the error with inverse $\chi$.
What is perhaps less obvious is whether approximate contraction trees with given $\chi$ should yield comparable errors regardless of the cost of the particular tree, $M$ or $C$. We see that this is in fact the case for the hyper-optimized trees, i.e.~the error correlates reasonably well with the compressed bond dimension $\chi$, independent of the choice of tree. Thus by choosing the optimized tree with lowest cost for a given $\chi$, we are not paying a price in terms of accuracy.

On the other hand, the hand-coded algorithms do not follow this observation, e.g. CTMRG in the 2D lattice, and boundary PEPS and HOTRG in the 3D lattice exhibit considerably larger errors than the hyper-optimized strategies for given $\chi$. For fixed bond dimension, the hyper-optimized contraction trees appear to use the computational resources (memory and cost) in a more effective way to reduce error than the hand-coded strategies.}

\begin{figure*}[t!]
    \centering
    \includegraphics[width=\textwidth]{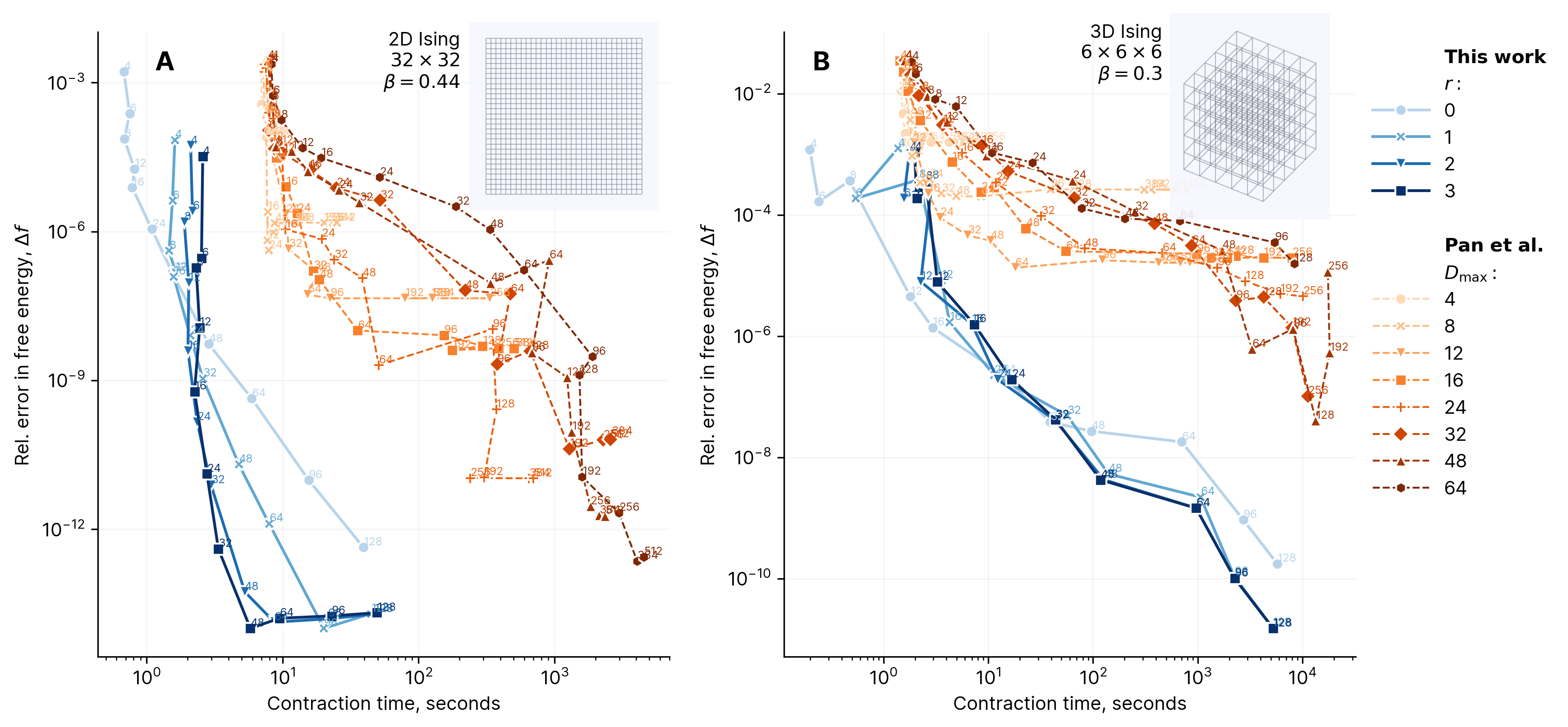}
    \caption{
        Performance comparison of hyper-optimized approximate contraction (current work) and the algorithm of Pan et al.~\cite{pan2020contracting} for computing the free energy of the Ising model at approximately the critical point of \textbf{A}: a square lattice and \textbf{B}: a cubic lattice.
        For both algorithms $\chi$ is varied and the points are labeled with the value.
        The insets show the geometry and specific sizes of the lattices.
    }\label{fig:performance-comparison}
\end{figure*}

\subsection{Error versus cost}

\REVISE{We next consider the error obtained for a given peak memory or computational cost.
In Fig.~\ref{fig:performance-ctmrg-hotrg-small} we show the relative error in the contraction value $\Delta Z$/free energy per site $\Delta f$ for 2D and 3D Ising and random tensor models for the hyper-optimized contraction and handcoded strategies, plotted against peak memory usage or contraction cost (depending on which was used as the cost function to optimize the trees). In this figure, the sizes of the problems were chosen so that the exact value of $Z$ or $f$ can be computed by exact TN contraction. In Fig.~\ref{fig:performance-ctmrg-hotrg-large} we consider the same models, but now for problem sizes too large for exact contraction. In these cases, we use $d\ln |Z|/d\chi$ and $df/d\chi$ as a metric of the convergence of the calculation.

From these plots we observe a few features. In the 2D models, the optimized \texttt{Span}, \texttt{Greedy}, and standard boundary contraction algorithms generally all achieve quite similar performance, and are the best performing algorithms. (We note that although \texttt{Span} shows a consistent advantage over boundary contraction in the error versus $\chi$ plots in Sec.~\ref{sec:err_vs_bond}, it does not do so when the  overall cost/memory is considered, because this depends on additional details besides $\chi$, such as the number of large tensors, order in which they contracted, etc.). CTMRG, HOTRG, and \texttt{Agglom} also perform similarly, and all perform much worse than the
\texttt{Span}, \texttt{Greedy}, and standard boundary contraction algorithms on this regular 2D latice.

In 3D, the PEPS boundary, CTMRG, and HOTRG all perform quite poorly, while \texttt{Span} performs well. \texttt{Greedy} performs well in the smaller examples, but degrades in the larger lattice, presumably again because of the limited contraction tree space generated by the \texttt{Greedy} algorithm. As noted in Sec.~\ref{sec:err_vs_bond}, hyper-optimized \texttt{Span} trees choose a quite different contraction path than the PEPS boundary algorithm, while still taking advantage of the boundary, and this is key to the improved performance.

Taken in total, the comparisons in the last three subsections illustrate how the optimized approximate contraction trees are competitive with, and can even exceed, the performance of standard contraction strategies in the simple lattices studied in many-body physics applications.}

\subsection{Comparison to another strategy for general graphs}

We next turn to a comparison of our hyper-optimized approximate contraction strategy to another recently proposed technique for arbitrary graphs.
Ref.~\cite{pan2020contracting}, proposed an algorithm to automatically contract arbitrary geometry tensor networks with a good performance across a range of graphs. For convenience we refer to that algorithm as \texttt{CATN}. \REVISE{
As we cannot trace through \texttt{CATN} in the same way as our previous performance comparisons, we measure the contraction time directly on a single CPU. Although \texttt{CATN} is formulated in a geometry independent manner, a critical difference with the current work is that}
\texttt{CATN} does not optimize over families of approximate contraction trees.

In Fig.~\ref{fig:performance-comparison} we compare
\texttt{CATN} against hyper-optimized \texttt{Span} trees for the 2D/3D Ising model at approximately the critical point, as a function of accuracy versus contraction time.
For the \texttt{Span} trees, we sweep over $\chi$ for different choices of tree gauge distance $r$ (i.e.~each line corresponds to one $r$, while $\chi$ is swept over). The performance of \texttt{CATN} is considered as a function of the
two bond dimensions $D_\text{max}$ and $\chi$ (each line corresponds to a given $D_\text{max}$, while $\chi$ is swept over).
\REVISE{
We do not include the time to find the tree for the hyper-optimized contraction since one generally re-uses this many times.
However as a rough guide, for the lattices in Fig.~\ref{fig:performance-comparison}
the search converges to a good tree in 10--20 seconds.
}
Further details and comparisons are in the SI~\cite{si}. We see clearly that in both the 2D and 3D cases (Figs.~\ref{fig:performance-comparison}A and B respectively) the hyper-optimized \texttt{Span} trees achieve a better accuracy versus contraction time trade-off than the \texttt{CATN} algorithm. \REVISE{Given that \texttt{CATN} itself has an ordering of compressions} an interesting question is to what extent the strategy of \texttt{CATN} might also be optimized.

\section{The power of hyper-optimized approximate contraction}\label{sec:power}

We now illustrate the power of the hyper-optimized approximate contraction protocol defined above in \REVISE{a further range of} interesting problems.

\begin{figure*}[t!]
  \centering
  \includegraphics[width=0.9\textwidth]{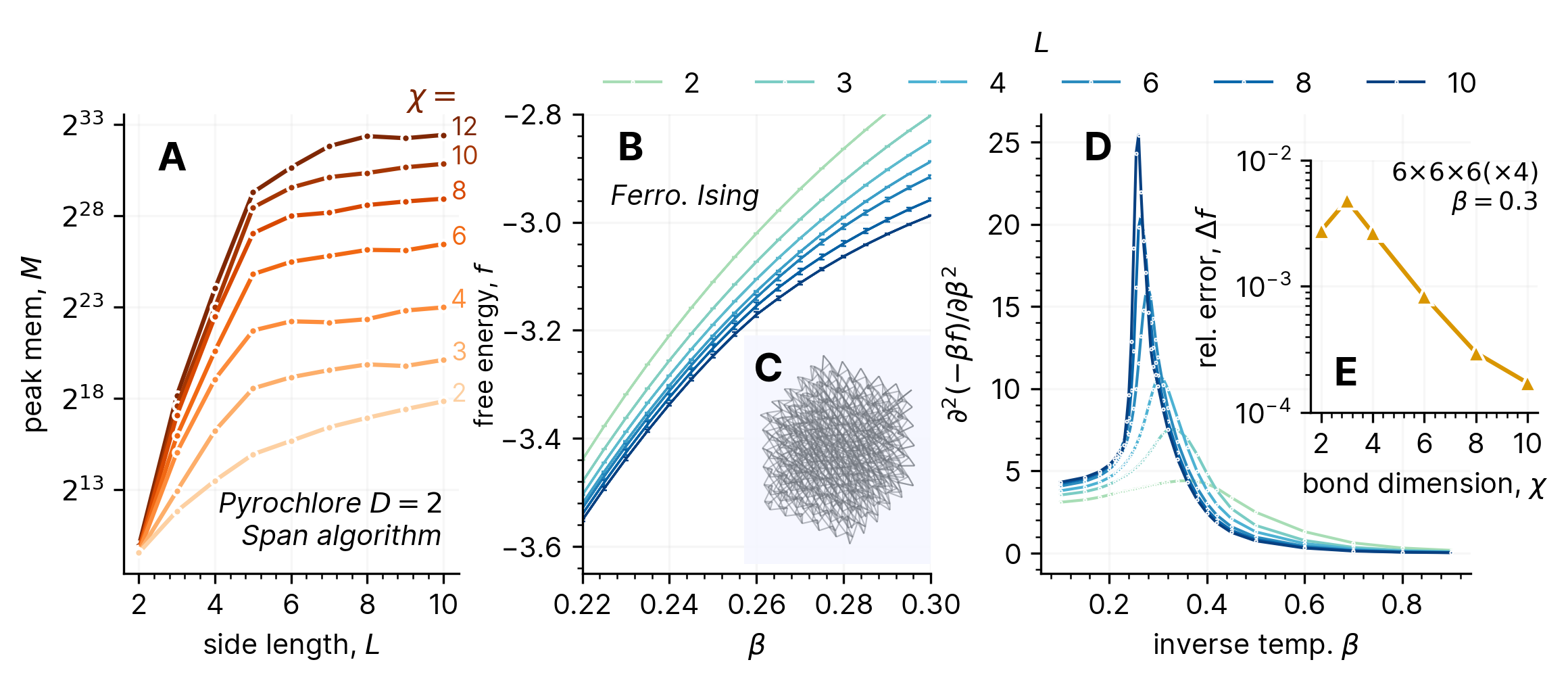}
  \caption{
  Approximate contraction of the ferromagnetic Ising model on the pyrochlore lattice.
  \textbf{A}: peak memory, $M$, of the \texttt{Span} algorithm on the pyrochlore lattice with $D{=}2$ as function of side length $L$ and $\chi$.
  \textbf{B}: the free energy, $f$, near the critical point, estimated by extrapolating approximate contractions in $\chi$. A tree-gauge distance of $r=2$ was used and the error bars show fit uncertainty.
  \textbf{C}: example instance of the pyrochlore geometry for $L=6$ corresponding to 864 sites.
  \textbf{D}: The second derivative of $-\beta f$ with respect to $\beta$, showing a diverging peak around the critical point for increasing $L$.
  \textbf{E}: relative error in free energy near the critical point as a function of $\chi$ for $L=6$ and $r=2$.
  }\label{fig:applications-pyrochlore}
 \end{figure*}

\subsection{Ising partition function on the pyrochlore  lattice}

We consider a tensor network contraction corresponding to the Ising partition function on large, finite, pyrochlore lattices \REVISE{with up to 4000 sites}. (We consider the version of the model where all spins are aligned along the \emph{same} axis, see Ref.~\cite{bramwell1998frustration}).
The highly frustrated geometry makes it harder to compute a low complexity contraction path for this tensor network than in simpler lattices.
In Fig.~\ref{fig:applications-pyrochlore}A we show the peak memory for the optimized \texttt{Span} algorithm as a function of side length $L$. The total lattice size is $L\times L \times L \times 4$, thus the largest calculation ($L=10$) is a contraction of 4000 tensors. For $L > 6$ we see the peak memory starts to saturate.
By fitting $f(\chi) = A + B / \chi$ to our data for parameters $A$ and $B$ we can accurately estimate both the free energy and its error as the fitted value and square root variance of the parameter $A=f(\infty)$.
We show the result of this in Fig.~\ref{fig:applications-pyrochlore}B, in the vicinity of the critical point~\cite{passosRepresentationSimulationPyrochlore2016,soldatovLargescaleCalculationFerromagnetic2017}.
We note that while Ising systems like this can be studied using Monte Carlo techniques, the partition function itself is tricky to estimate, requiring methods~\cite{wangEfficientMultipleRangeRandom2001} beyond the standard Metropolis algorithm~\cite{metropolisEquationStateCalculations1953}.
Fig.~\ref{fig:applications-pyrochlore}D shows the second derivative of $\frac{1}{N}\ln Z$  with respect to inverse temperature: ${\partial^2(-\beta f)}/{\partial \beta^2}$,
which displays a growing peak as a function of system length $L$, illustrating the critical point.

The largest \emph{exactly} contractable tensor network corresponds to size $L=6$; it is visualized in the inset Fig.~\ref{fig:applications-pyrochlore}C.
For this size we can investigate the free energy error of the approximate contraction scheme, $\Delta f$, and this is shown in Fig.~\ref{fig:applications-pyrochlore}E.
We see that increasing $\chi$ reliably decreases the error, and
for the largest $\chi$ considered, the relative free energy is only $10^{-4}$.

\subsection{Random 3-regular graphs and dimer coverings}

\begin{figure*}[t!]
  \centering
  \includegraphics[width=0.8\textwidth]{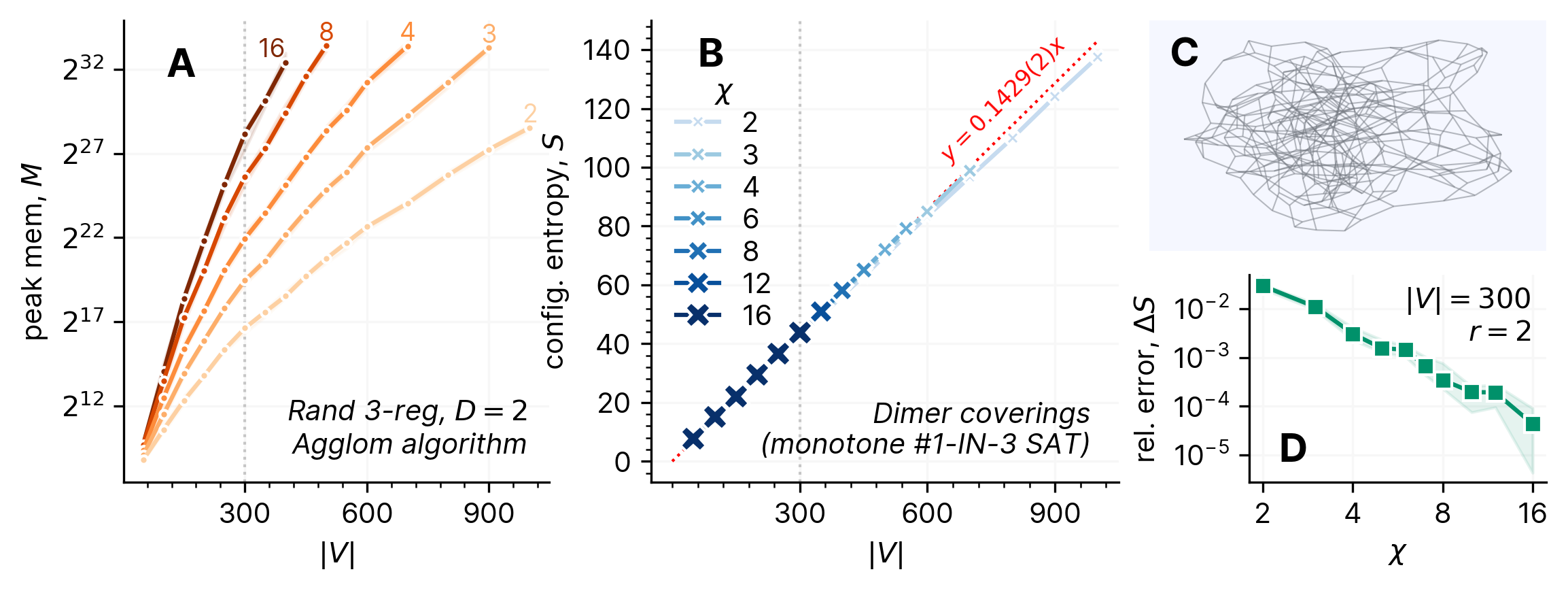}
  \caption{Approximate contraction of dimer covering counting on random 3-regular graphs. Quantities are averaged over 20 instances.
  \textbf{A}: peak memory, $M$, of the \texttt{Agglom} algorithm on random 3-regular graph instances with $D=2$ as a function of number of vertices, $|V|$.
  \textbf{B}: configuration entropy, $S=\ln W$, where $W$ is the number of valid configurations as a function of $|V|$ and $\chi$. The red dotted line shows the constant from a least squares fit to a quadratic function of inverse $|V|$ and $\chi$ (see main text).
  \textbf{C}: example random regular graph for $|V|=300$.
  \textbf{D}: relative error in $S$ as a function of $\chi$ for $|V|=300$.
  }\label{fig:applications-randreg}
\end{figure*}

We next study a problem defined on random 3-regular graphs. Here the \texttt{Agglom} algorithm performs best, and we show the resulting complexities in terms of peak memory, $M$, in Fig.~\ref{fig:applications-randreg}A.
Here, because the length of loops in the graph grows with increasing number of vertices, $|V|$, we still find an exponential scaling, even when compressing to fixed $\chi$.
Nonetheless, we can usefully push much beyond exactly contractible limit of $|V|\sim 300$ (illustrated in Fig.~\ref{fig:applications-randreg}C).
To study the accuracy we consider the problem of counting dimer coverings on these graphs.
This is equivalent to so-called positive \#\textsc{1-in-3SAT}~\cite{raymondPhaseDiagram1in32007,zdeborovaConstraintSatisfactionProblems2008,kourtis2019fast}.
Each edge (i.e.~index) is considered a potential dimer, and by placing the tensor
\begin{equation}
    T_{i, j, k} =
    \begin{cases}
    1, & \text{if} \ i + j + k = 1 \\
    0, & \text{otherwise}
    \end{cases}
\end{equation}
on each vertex, we enforce that every tensor be ``covered'' by a single dimer only for any configuration to be valid.
The decision version of this problem is NP-Complete~\cite{schaeferComplexitySatisfiabilityProblems1978,gareyComputersIntractabilityGuide1979}, and on random 3-regular graphs specifically the problem is known to be close, but just on the satisfiable side, in terms of ratio of clauses to variables, of the hardest regime~\cite{raymondPhaseDiagram1in32007,zdeborovaConstraintSatisfactionProblems2008}.
The contraction of the above tensor network gives the number of configurations, $Z$, at zero-temperature, and a corresponding `residual' entropy, $S=\ln Z$.
We plot $S$ in Fig.~\ref{fig:applications-randreg}B.
Considering the entropy per site, $s/|V|$, by performing a least squares fit with a quadratic function of inverse size, $1/|V|$ and bond dimension, $1/\chi$,
\begin{align}
  s(|V|, \chi) =
  s_{\infty} +
  \dfrac{c_1}{|V|^{2}} +
  \dfrac{c_2}{|V|}+
  \dfrac{c_3}{|V| \chi} +
  \dfrac{c_4}{\chi}+
  \dfrac{c_5}{\chi^{2}}
\end{align}
with fitted parameters, $s_{\infty}, c_{1\ldots 5}$,
we can estimate the infinite size entropy per site as $s_{\infty}=0.1429(2)$.
The theoretical value of this can be computed using~\cite{bollobas1986number} when $|V| \gtrsim 5{\times}10^{11}$ (see the SI), yielding 0.1438, suggesting a small systematic error remains from the finite size.
In Fig.~\ref{fig:applications-randreg}D we consider the relative error in $S$ when compared to exact contraction results for $|V|=300$, where again we see that increasing $\chi$ reliably improves the error.

\subsection{Hardness transition in random tensor networks}

\begin{figure*}[t!]
  \centering
  \includegraphics[width=0.8\textwidth]{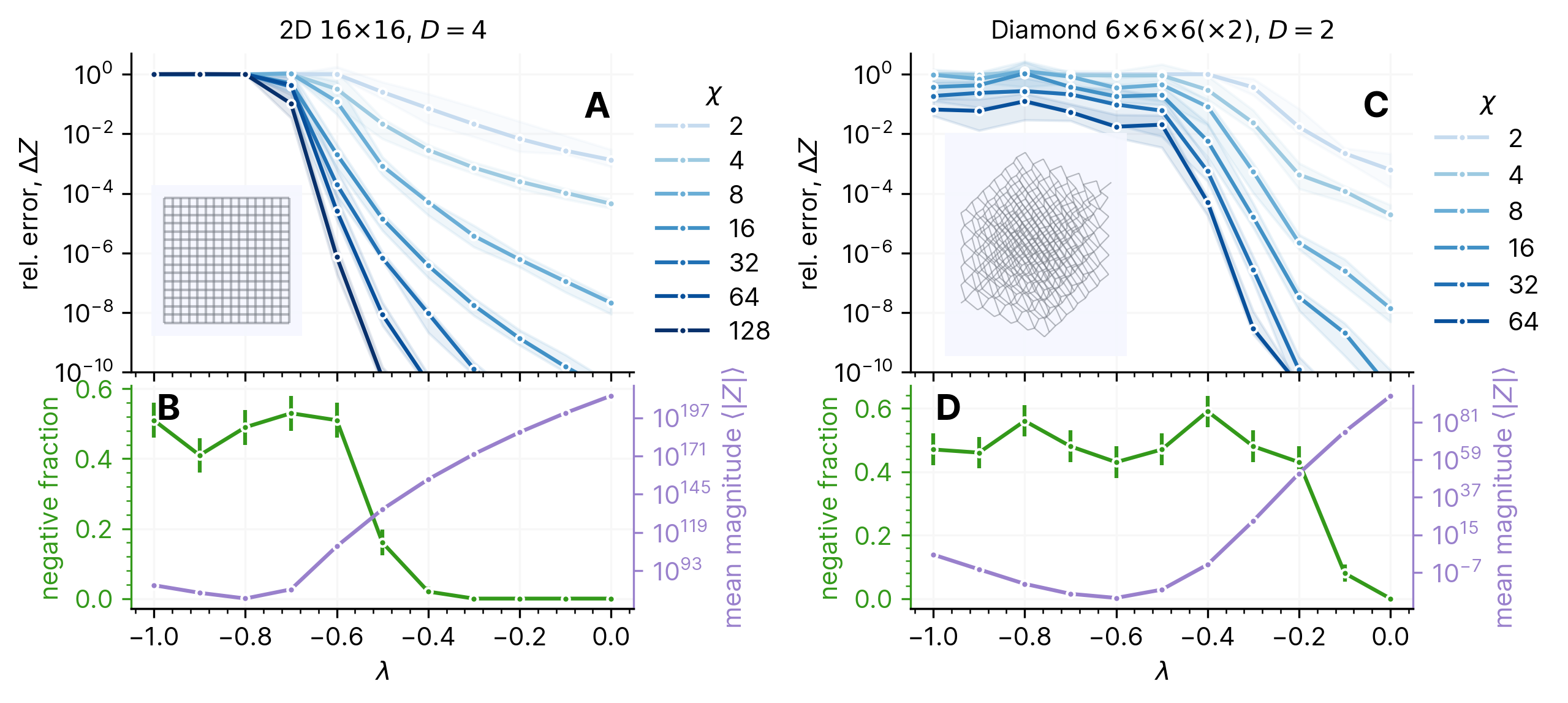}
  \caption{Hardness transition in approximately contracting tensor networks with random uniform entries $\in [\lambda, 1]$.
  \textbf{A}: relative error, $\Delta Z$, in approximately contracted value of the URand model on the square lattice using the \texttt{Greedy} algorithm as a function of $\lambda$ and $\chi$ with $r=2$. Line and bands show median and interquartile range across 20 instances.
  \textbf{B}: distribution of actual values $Z$ for the square URand model in terms of fraction of negative instances (green, left axis) and average absolute magnitude (purple, right axis). Error bars denote error on mean.
  \textbf{C}: relative error, $\Delta Z$, in approximately contracted value of the URand model on the diamond lattice using the \texttt{Greedy} algorithm as a function of $\lambda$ and $\chi$ with $r=2$. Line and bands show median and interquartile range across 20 instances.
  \textbf{D}: distribution of actual values $Z$ for the diamond URand model in terms of fraction of negative instances (green, left axis) and average absolute magnitude (purple, right axis). Error bars denote error on mean.
  }\label{fig:applications-hardness}
\end{figure*}

The final problem we consider is one where the hardness derives from the tensor entries themselves rather than the geometry.
We take the URand model -- with tensor entries sampled uniformly $\in [\lambda, 1]$ -- and consider two lattices which are amenable to contraction with relatively large $\chi$, the square and diamond lattices.
We take sizes $16{\times}16$ with $D=4$ and $6{\times}6{\times}6({\times}2)$ with $D=2$ respectively, both of which are at the limit of what is contractible exactly.
In Fig.~\ref{fig:applications-hardness}A we show the relative error in the approximately contracted value, $\Delta Z$, as a function of $\lambda$ across 20 random instances.
There is a clear transition in hardness at $\lambda \sim -0.7$ -- above this even moderate $\chi$ is sufficient to contract the tensor network with very good accuracy.
Below this however, there is no improvement to the error at all with increasing $\chi$; the contracted value remains essentially impossible to approximate.
An obvious question is how does $Z$ itself change with $\lambda$?
In Fig.~\ref{fig:applications-hardness}B we show the fraction of instances whose exact value $Z$ is negative, as well as the average magnitude of $Z$.
The problem varies from smaller magnitude values (compared to the total number of terms in the sum, $\sim 10^{300}$) evenly split between negative and positive, to large always positive values.
We also consider the same model but embedded in a 3D diamond geometry in Fig.~\ref{fig:applications-hardness}C.
The same transition in hardness occurs at a slightly different value of $\lambda$. In the hard regime, there remains some small ability to approximate $Z$ with large $\chi$, probably as this is approaching exact contraction. The complexity of contraction of the random tensor network is thus closely related to the positive nature of the tensor entries. This is likely related to the conjectured low entanglement of typical positive tensor networks~\cite{groverEntanglementSignStructure2015}, as well as
the hardness of approximating complex valued Ising models~\cite{goldbergComplexityApproximatingComplexvalued2017,galanisComplexityApproximatingComplexvalued2021,buysLeeYangZeros2022}.

\section{Conclusions}

We have introduced a framework for approximate contractions of  tensor networks defined on arbitrary graphs, based on hyper-optimizing over ordered contraction trees \REVISE{with compression steps.} In particular, our work attempts to optimize over the many choices and components in such an algorithm,
ranging from the manner in which compressions are performed, to the sequence and ordering of compression and contractions.
\REVISE{Interestingly, we observe that by minimizing a cost function associated with memory or computational cost, we simultaneously generate an approximate contraction tree that yields small contraction error.}
In many cases, we find that the optimization produces significantly cheaper and more accurate contraction strategies than hand crafted approximate contraction algorithms, even in well studied regular lattices.
While we cannot claim that our final algorithm is optimal, the purpose of the framework is to allow an optimization over compression strategies, and such an optimization can be extended should, for example, other metrics of approximate contraction quality be introduced. We envisage that the many constituent parts of our protocol can be separately improved in future works.

We have discussed different regimes of computational advantage for approximate contraction over exact contraction.
Firstly, for locally connected graphs and ``non-hard'' tensor entries, we expect approximate contraction to display an exponential benefit over exact contraction, as shown here for the pyrochlore lattice. Secondly, for certain geometries with long range interactions, we expect approximate contraction to still scale exponentially, but with a usefully reduced pre-factor, as shown here for 3-regular random graphs. Finally, we expect some classes of tensor entries to be essentially impossible to approximately contract, regardless of geometry, as shown here for certain distributions of random tensors. Whether the latter result corresponds to the hardness of contraction expected for generic random quantum circuits, is an interesting question. Similarly, the application of these techniques to quantum circuit and ansatz expectation values is a natural direction that we leave for future work.

\begin{acknowledgments}
We thank Stefanos Kourtis and Pan Zhang for helpful comments on this manuscript, and Hitesh Changlani for help with understanding the pyrochlore lattice.

The development of the contraction tree optimization algorithms was supported by the US National Science Foundation through Award No. 1931328. The approximate contraction algorithms were developed with support from the US National Science Foundation through Award No. 2102505.  JG acknowledges support through a gift from Amazon Web Services Inc. GKC is partially supported as a Simons Investigator in Physics.
\end{acknowledgments}

\clearpage
\onecolumngrid
\begin{center}
    \large \textbf{
    Hyper-optimized approximate contraction of tensor networks with arbitrary geometry: supplementary information
    }
\end{center}
\appendix
\tableofcontents
\newpage

\section{Tree spans and gauging}\label{sec:tree-spans}

In this section we detail the simple method of generating a (possibly $r$-local) spanning tree, for use with the tree gauge, and also in the \texttt{Span} contraction tree building algorithm (note the spanning tree is a different object to the contraction tree).
A pseudocode outline is given in Algorithm.~\ref{alg:local-spanning-tree}.
We first take an arbitrary initial connected subgraph, $S_0$ of the graph, $G$, for example the two tensors sharing bonds that are about to be compressed.
We then greedily select a pair of nodes, one inside and one outside this region, which expands the spanning tree, $\tau$ and region $S$, until all nodes within graph distance $r$ of $S_0$ are in $S$.
Since the nodes can share multiple edges, the spanning tree $\tau$ is best described using an ordered set of node pairs rather than edges.
If the graph $G$ has cycles, then nodes outside $S$ may have multiple connections to it, this degeneracy is broken by choosing a scoring function.

\begin{algorithm}[H]
\caption{$r$-local spanning tree}\label{alg:local-spanning-tree}
\begin{algorithmic}
\State \textbf{Input:}  graph $G$, initial region $S_0$, max distance $r$
\State $\tau \gets \{\}$ \Comment{ordered set of pairs forming spanning tree}
\State $S \gets S_0$ \Comment{set of nodes spanned by the tree}
\State $c \gets \{\}$ \Comment{candidates to add to tree}
\For{$u \in S$} \Comment{each node in original region}
\For{$v \in \textsc{neighbors}(G, u) \setminus S$} \Comment{connected nodes not in region}
\State $r_{uv} \gets 1$ \Comment{distance to original region}
\State $c \gets c \cup \{(u, v, r_{uv})\}$
\EndFor
\EndFor
\While{$|c| > 0$}
\State $u, v, r_{uv} \gets \textsc{best}(c)$ \Comment{pop the best candidate edge}
\State $c \gets c \setminus \{(u, v, r_{uv})\}$
\If{$(v \notin S) \wedge (r_{uv} \leq r)$ } \Comment{node is new and close enough}
\State $S \gets S \cup \{v\}$ \Comment{add $v$ to region}
\State $\tau \gets \tau \cup \{(u, v)\}$ \Comment{add edge to tree}
\For{$w \in \textsc{neighbors}(G, v) \setminus S$} \Comment{add new neighboring candidates}
\State $r_{vw} \gets r_{uv} + 1$
\State $c \gets c \cup \{(v, w, r_{vw})\}$
\EndFor
\EndIf
\EndWhile
\State \textbf{Return:} $\tau$, $S$
\end{algorithmic}
\end{algorithm}

For the tree gauge, we choose the scoring function such that the closest node with the highest connectivity (product of sizes of connecting edge dimensions) is preferred.
The gauging proceeds by taking the pairs in $\tau$ in reverse order, gauging bonds from the outer to the inner tensor (see main text Fig.3E).
If the graph $G$ is a tree and we take $r=\infty$, this corresponds exactly to canonicalization of the region $S_0$.
In order to perform a compression of a bond in the tree gauge, we just need to perform QR decompositions inwards on a `virtual copy' of the tree (as shown in the main text Fig.~4B), until we have the central 'reduced factors' $R_A$ and $R_B$. Performing a truncated SVD on the contraction of these two to yield $R_A R_B = R_{AB} \approx U, \sigma, V^\dagger$, allows us to compute the locally optimal projectors to insert on the bond as $P_L= R_B V \sigma^{-1/2}$ and $P_R = \sigma^{-1/2} U^\dagger R_A$ such that $AB\approx A P_L P_R B$.
The form of these projectors, which is the same as CTMRG and HOTRG but including information up to distance $r$ away, is explicitly derived in Sec.~\ref{sec:projectors}.
One further restriction we place is to exclude any tensors from the span that are input rather than intermediate tensors.

One obvious alternative possibility to the tree-gauge is to introduce an initial `Simple Update' style gauge~\cite{jiangAccurateDeterminationTensor2008} on each of the bonds and update these after compressing a bond, including in the vicinity of the adjacent tensors.
A similar scheme was employed for 3D contractions in \cite{VlaarSimulationthreedimensionalquantum2021}.
In our experience this performs similarly to the tree-gauge (and indeed the underlying operations are very similar) but is more susceptible to numerical issues due to the direct inversion of potentially small singular values.

\section{Explicit projector form} \label{sec:projectors}

\begin{figure}[t!]
    \centering
    \includegraphics[width=\linewidth]{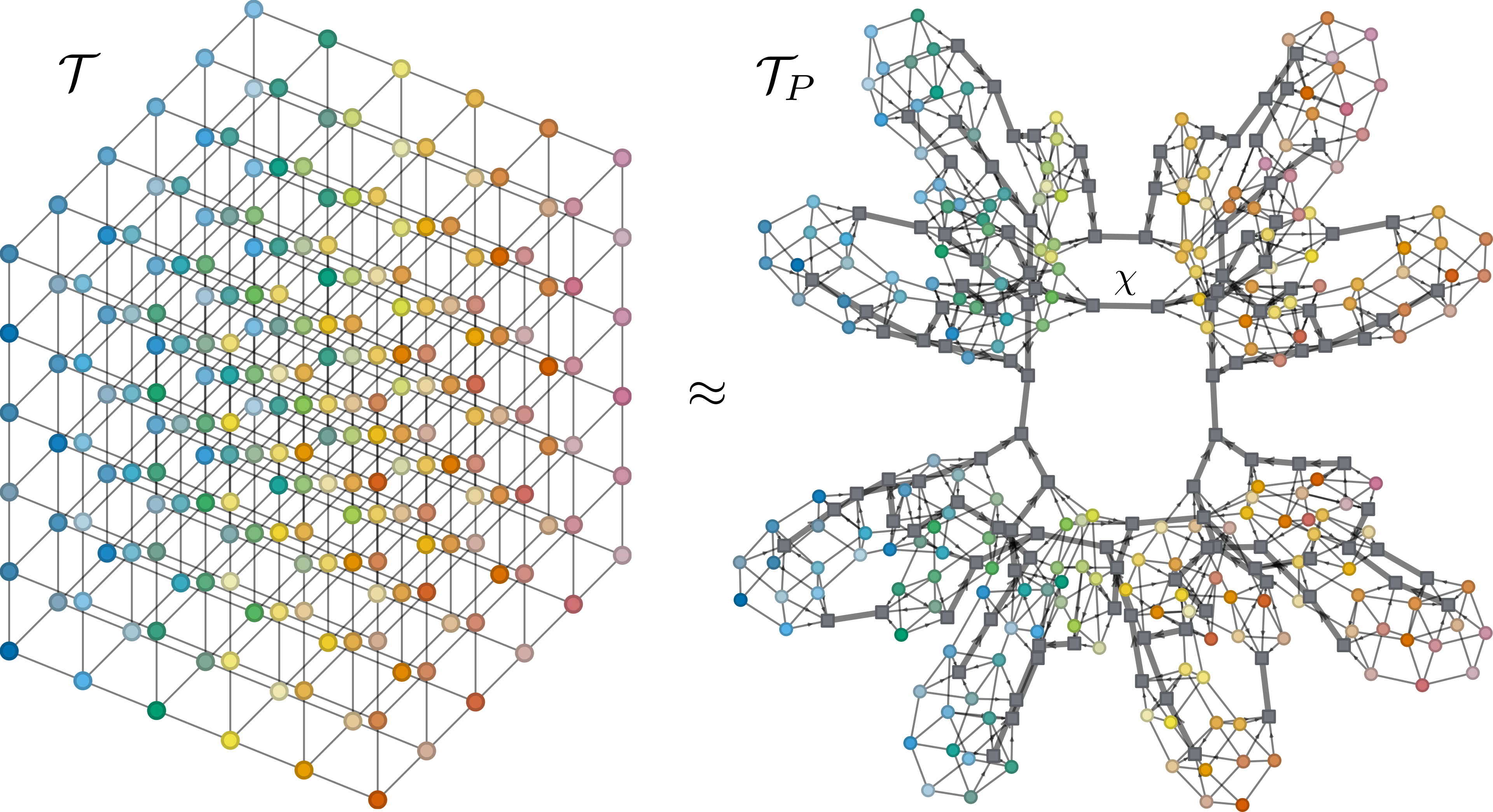}
    \caption{An example of transforming a tensor network, $\mathcal{T}$, into an exactly contractable tensor network, $\mathcal{T}_P$, using the explicit projector form of a approximate contraction tree.
    Here we take a $6{\times}6{\times}6$ with $D=2$ cube and use an optimized contraction tree from the $\texttt{Span}$ generator, for $\chi=8$.
    Each node in the original tensor network is colored uniquely. The grey square nodes in the right hand side diagram represent the inserted projectors, with thicker edges the compressed bonds of size $\chi$.
    Arrows indicate the orientation of the projectors (i.e. the order of the compressions).
    }
    \label{fig:explicit-projection}
\end{figure}
Performing a bond compression such as in the main text Fig.~3D can be equated to the insertion  of two approximate projectors that truncate the target bond to size $\chi$. The projector form allows us to perform the tree gauge compression 'virtually' - i.e. without having to modify tensors anywhere else in the original tensor network. We begin by considering the product $AB$, where $A$ and $B$ might represent collections of tensors such as a local tree. Assuming we can decompose each into a orthogonal and `reduced' factor we write:
\begin{align*}
    AB = (Q_A R_A)(R_B Q_B)~.
\end{align*}
If we resolve the identity on either side, we can form the product $R_A R_B$ in the middle and perform a truncated SVD on this combined reduced factor yielding $U \sigma V^{\dagger}$.
\begin{align*}
    AB &= Q_A R_A (R_A^{-1} R_A) (R_B R_B^{-1}) R_B Q_B \\
    &= Q_A R_A R_A^{-1} (U \sigma V^\dagger) R_B^{-1} R_B Q_B \\
    &= Q_A R_A (R_A^{-1} U \sqrt{\sigma}) (\sqrt{\sigma} V^\dagger R_B^{-1}) R_B Q_B
\end{align*}
from which we can read off the projectors that we need to insert into the original tensor network in order to realize the optimal truncation as:
\begin{align*}
    P_L &= R_A^{-1} U \sqrt{\sigma}~, \\
    P_R &= \sqrt{\sigma} V^\dagger R_B^{-1} ~.
\end{align*}
Finally, in order to avoid performing the inversion of the reduced factors, we can simplify:
\begin{align*}
    P_L &= R_A^{-1} U \sqrt{\sigma} \\
        &= R_A^{-1} (U \sqrt{\sigma} \sqrt{\sigma} V^\dagger) V \sigma^{-1/2} \\
        &= R_A^{-1} (R_A R_B) V \sigma^{-1/2} \\
        &= R_B V \sigma^{-1/2}
\end{align*}
and likewise:
\begin{align*}
    P_R &= \sqrt{\sigma} V^\dagger R_B^{-1} \\
        &= \sigma^{-1/2} U^\dagger (U \sqrt{\sigma} \sqrt{\sigma} V^\dagger) R_B^{-1} \\
        &= \sigma^{-1/2} U^\dagger (R_A R_B) R_B^{-1} \\
        &= \sigma^{-1/2} U^\dagger R_A~.
\end{align*}
This form of the projectors makes explicit the equivalence to CTMRG and HOTRG~\cite{wangClusterUpdateTensor2011,corbozCompetingStatesTJ2014,iinoBoundaryTensorRenormalization2019}, for which $R_A$ and $R_B$ contain only information about the local plaquette. Note in general that we just need to know $R_A$ and $R_B$ (not $Q_A$ or $Q_B$) to compute $P_L$ and $P_R$, but we can include in these the effects of the distance-$r$ tree gauge in order to perform the truncation locally without modifying any tensors but $A$ and $B$.

Rather than dynamically performing the approximate contraction algorithm using the ordered contraction tree, one can also use it to statically map the original tensor network, $\mathcal{T}$, to another tensor network, $\mathcal{T}_P$, which has the sequence of projectors lazily inserted into it (i.e. each $A P_L P_R B$ is left uncontracted).
\emph{Exact} contraction of $\mathcal{T}_P$ then gives the approximate contracted value of $\mathcal{T}$.
Such a mapping may be useful for relating the approximate contraction to other tensor network forms~\cite{ranLectureNotesTensor2020}, or for performing some operations such as optimization~\cite{xieSecondRenormalizationTensorNetwork2009}.
Here we describe the process.

To understand where the projectors should be inserted we just need to consider the sub-graphs that the intermediate tensors correspond to.
At the beginning of the contraction, each node corresponds to a sub-graph of size 1, containing only itself. We can define the sub-graph map $S(i) = \{i\}$ for $i = 1 \ldots N$.
When we contract two nodes $i, j$ to form a new node $k$, the new sub-graph is simply $S(k) = S(i) \cup S(j)$. When we compress between two intermediate tensors $i$ and $j$, we find all bonds connecting $S(i)$ to $S(j)$, and insert the projectors $P_L$ and $P_R$, effectively replacing the identity linking the two regions with the rank-$\chi$ operator $P_L P_R$.
Finally we add the tensor $P_L$ to the sub-graph $S(i)$ and $P_R$ to the sub-graph $S(j)$.
This can be visualized like so.
\begin{center}
\includegraphics[width=0.7\linewidth]{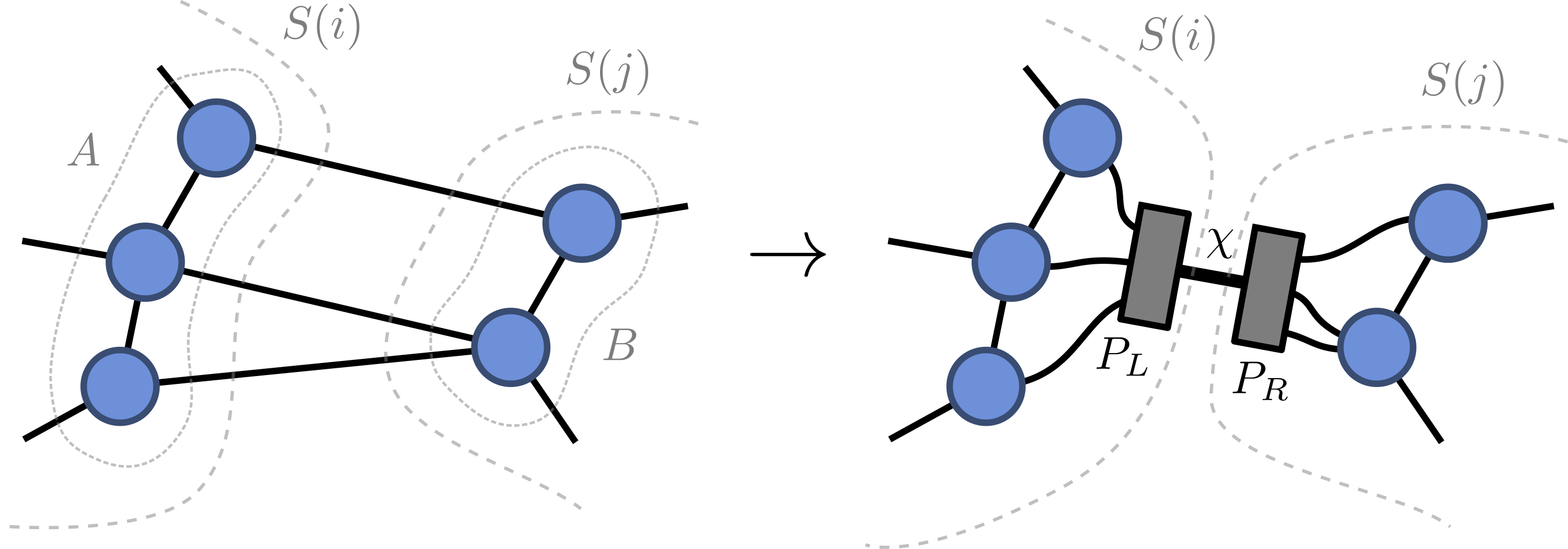}
\end{center}
Grouping all the neighboring tensors on one side of the bonds as an effective matrix $A$ and those on the other side as $B$ (note that these might generally include projectors from previous steps), the form of $P_L$ and $P_R$ can be computed as above.

An example of the overall geometry change of performing this explicit projection transformation for the full set of compressions on a cubic tensor network approximate contraction is shown in Fig.~\ref{fig:explicit-projection}.
Note that the dynamic nature of the projectors, which depend on both the input tensors and the contraction tree, is what differentiates a tensor network which you contract using approximate contraction, and for instance directly using a tree- or fractal-like ansatz such as $\mathcal{T}_P$.

\section{Tree builder details}\label{sec:tree-builders}

In this section we provide extended details of each of the heuristic ordered contraction tree generators.
First we outline the hyper optimization approach.
Each tree builder $B$ takes as input the graph $G$ with edges weighted according to the tensor network bond sizes, as well as a set of heuristic hyper-parameters, $\bar\theta$, that control how it generates an ordered contraction tree $\Upsilon$.
The builder is run inside a hyper-optimization loop that uses a generic optimizer, $O$, to sample and tune the parameters.
We use the \texttt{nevergrad}~\cite{nevergrad} optimizer for this purpose.
A scoring function computes some metric $y$ for each tree (see Sec.~\ref{sec:metrics} for possible functions), which is used to train the optimizer and track the best score and tree sampled so far, $y_{best}$ and $\Upsilon_{best}$ respectively.
The result, outlined in Algorithm~\ref{alg:hyperopt}, is an anytime algorithm (i.e. can be terminated at any point) that samples trees from a space that progressively improves.
Note that while the optimization targets a specific $\chi$, the tree produced exists separately from $\chi$ and can be used for a range of values of $\chi$ (in which case one would likely optimize for the maximum value).

\begin{algorithm}[H]
\caption{Hyper optimization loop}\label{alg:hyperopt}
\begin{algorithmic}
\State{\textbf{Input:}  graph $G$, max bond $\chi$, builder $B$, optimizer $O$}
\State $y_{best} \gets \infty$
\While{$optimizing$}
\State $\bar{\theta} \gets \textsc{sample\_parameters}(O)$ \Comment{Get new hyper parameters}
\State $\Upsilon \gets \textsc{generate\_tree}(B, G, \bar{\theta})$ \Comment{Build tree with new parameters}
\State $y \gets \textsc{score\_tree}(\Upsilon, \chi)$ \Comment{Score the tree}
\If {$y < y_{best}$}
\State $y_{best} \gets y$
\State $\Upsilon_{best} \gets \Upsilon$
\EndIf
\State $\textsc{report\_parameters}(O, \bar{\theta}, y)$ \Comment{Update optimizer with score}
\EndWhile
\State \textbf{Return:} $\Upsilon_{best}$
\end{algorithmic}
\end{algorithm}

In the following subsections we outline the specific hyper parameter choices, $\bar\theta$, for each tree builder.
However one useful recurring quantity is a measure of \emph{centrality}, similar to the harmonic closeness\cite{beauchamp1965improved,marchiori2000harmony}, that assigns to each node a value according to how central it is in the network.
This can be computed very efficiently as $c^{[v]} = \frac{1}{Z} \sum_{u \neq v} \frac{1}{\sqrt{d(u, v) + 1}}$, where $d(u, v)$ is the shortest distance between nodes $u$ and $v$. The normalization constant $Z$ is chosen such that $c^{[v]}\in[0, 1]~\forall~v$.

\subsection{\texttt{Greedy}}\label{sec:builders-greedy}

The \texttt{Greedy} algorithm builds an ordered contraction tree by taking the graph at step $\alpha$ of the contraction, $G_{\alpha}$, and greedily selecting a pair of tensors to contract $(i, j) \rightarrow k$, simulating the contraction and compression of those tensors, and then repeating the process with the newly updated graph, $G_{\alpha + 1}$, until only a single tensor remains.
The pair of tensors chosen at each step are those that minimize a local scoring function, and it is the parameters within this that are hyper-optimized.
The local score is a sum of the following components:
\begin{itemize}
    \item $\log_2$ size of new tensor \emph{after} compression with weight $\theta_{new\_size}$.
    \item $\log_2$ size of new tensor \emph{before} compression with weight $\theta_{old\_size}$.
    \item The minimum, maximum, sum, mean or difference (the choice of which is a hyper parameter) of the two input tensor sizes $\log_2$, with weight $\theta_{inputs}$.
    \item The minimum, maximum, sum, mean or difference (the choice of which is a hyper parameter) of the sub-graph sizes of each input (when viewed as sub-trees) with weight $\theta_{subgraph}$.
    \item The minimum, maximum, mean or difference (the choice of which is a hyper parameter) of the centralities of each input tensor with weight $\theta_{centrality}$. Centrality is propagated to newly contracted nodes as the minimum, maximum or average of inputs (the choice of which is a hyper-parameter).
    \item a random variable sampled from the Gumbel distribution multiplied by a temperature (which is a hyper-parameter).
\end{itemize}

The final hyper-parameter is a value of $\chi_{greedy}$ to simulate the contraction with, which can thus deviate from the real value of $\chi$ used to finally score the tree.
The overall space defined is 11-dimensional, which is small enough to be tuned by, for example, Bayesian optimization. In our experience it is not crucial to understand how each hyper-parameter affects the tree generated, other than that they are each chosen to carry some meaningful information from which the optimizer can conjure a local contraction strategy; the approach is more in the spirit of high-dimensional learning rather than a physics-inspired optimization.

\subsection{\texttt{Span}}\label{sec:builders-span}

The \texttt{Span} algorithm builds an ordered contraction tree using a modified, tunable version of the spanning tree generator in Algorithm~\ref{alg:local-spanning-tree} with $r=\infty$. The basic idea is to interpret the ordered sequence of node pairs in the spanning tree, $\tau$, as the reversed series of contractions to perform.
The initial region $S_0$ is taken as one of the nodes with the highest or lowest centrality (the choice being a hyper-parameter).
The remaining hyper-parameters are used to tune the local scoring function ($\textsc{best}(c)$ in Algorithm.~\ref{alg:local-spanning-tree}), that decides which pair of nodes should be added to the tree at each step.
These are:
\begin{itemize}
    \item The connectivity of the candidate node to the current region, with weight $\theta_{connectivity}$.
    \item The dimensionality of the candidate tensor, with weight $\theta_{ndim}$.
    \item The distance of the candidate node from the initial region, with weight $\theta_{distance}$
    \item The centrality of the candidate node, with weight $\theta_{centrality}$
    \item a random variable sampled from the Gumbel distribution multiplied by a temperature (which is a hyper-parameter).
\end{itemize}
The final hyper-parameter is a permutation controlling which of these scores to prioritize over others.

\subsection{\texttt{Agglom}}\label{sec:builders-agglom}

The \texttt{Agglom} algorithm builds the contraction tree by repeated graph partitioning using the library \texttt{KaHyPar}~\cite{kahypar1,kahypar2}.
We first partition the graph, $G$ into $\sim |V| / K$ parts, with the target subgraph size $K$ being a tunable hyper-parameter.
Another hyper-parameter is the imbalance, $\theta_{imbalance}$, which controls how much the sub-graph sizes are allowed to deviate from $K$.
Other hyper-parameters at this stage pertain to \texttt{KaHyPar}:
\begin{itemize}
    \item $\theta_{mode}$ either `direct' or `recursive',
    \item $\theta_{objective}$ either `cut' or `km1',
    \item $\theta_{weight}$ whether to weight the edges constantly or logarithmically according to bond size.
\end{itemize}
Once a partition has been formed, the graph is transformed by simulating contracting all of the tensors in each group, and then compressing between the new intermediates to create a new graph with $\sim |V| / K$ nodes and bonds of size no more than $\chi_{agglom}$ (itself a hyper-parameter which can deviate from the real $\chi$ used to score the tree). The contractions within each partition are chosen according to the \texttt{Greedy} algorithm.
Finally, the tree generated in this way is not ordered. To fix an ordering the contractions are sorted by sub-graph size and average centrality.

\subsection{Branch \& bound approximate contraction tree} \label{sec:builders-bandb}

The hyper-optimized approach produces heavily optimized trees but with no guarantee that they are an optimal solution.
For small graphs a depth first branch and bound approach can be used to find an optimal tree exhaustively, or to refine an existing tree if terminated early.
The general idea is to run the greedy algorithm whilst tracking a score, but keep and explore every candidate contraction at each step (a `branch') in order to `rewind' and improve it.
The depth first aspect refers to prioritizing exploring branches to completion so as to establish an upper bound on the score.
The upper bound can then be improved and used to terminate bad branches early.

\begin{figure}
\begin{algorithm}[H]
  \caption{Branch and bound tree search}
  \label{alg:optimal-tree}
   \begin{algorithmic}[0]
   \State \textbf{Input:}  graph $G$, Maximum bond dimension $\chi$
   \State $y_{best} \gets \infty$
   \State c = \{\} \Comment{candidate contractions}
   \For{$i, j \in \textsc{edges}(G)$} \Comment{populate with every pair of tensors}
   \State $y \gets 0$ \Comment{initial score}
   \State $p \gets [~]$ \Comment{the contraction `path'}
   \State $c \gets c \cup \{(i, j, G, y, p)\}$
   \EndFor
   \While{$|c| > 0$}
   \State $(i, j, G, y, p) \gets \textsc{remove\_best}(c)$
   \If{$\textsc{invalid}(i, j, G)$ \textbf{or}  $y \geq y_{best}$} \Comment{no need to explore further}
   \State \textbf{continue}
   \EndIf
   \If{$|G|=1$ \textbf{and} $y < y_{best}$} \Comment{finished contraction with best score}
   \State $y_{best} \gets y$
   \State $p_{best} \gets p$
   \State \textbf{continue}
   \EndIf
    \State $p \gets \textsc{append}(p, (i, j))$ \Comment{continue exploring}
    \State $(k, G, y) \gets \textsc{simulate\_contraction}(i, j, G, \chi)$ \Comment{$k$ is the new node}
    \For{$l \in \textsc{neighbors}(G, k)$} \Comment{add new possible contractions}
    \State $c \gets c \cup \{(k, l, G, y, p)\}$
    \EndFor
   \EndWhile
   \State $\Upsilon_{best} \gets \textsc{build\_tree\_from\_path}(G, p_{best})$
   \State \textbf{Return:} $\Upsilon_{best}$
   \end{algorithmic}
\end{algorithm}
\end{figure}

\section{Tree cost functions} \label{sec:metrics}

\begin{figure}[htb]
    \centering
    \includegraphics[width=0.9\textwidth]{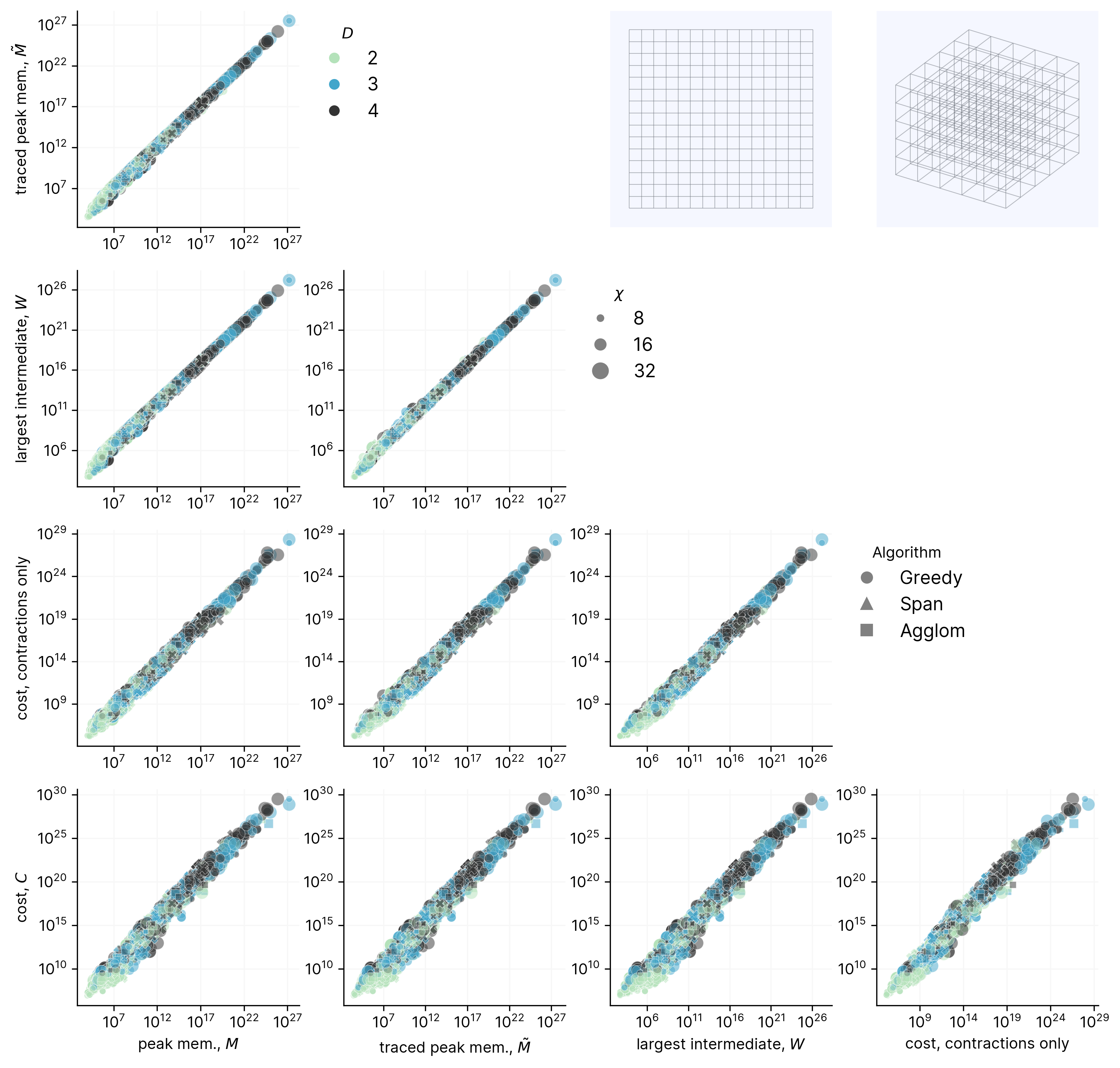}
    \caption{Relationship between various tree cost functions for randomly sampled approximate contraction trees for two geometries: 2D square of size $16\times16$ and 3D cube of size $6\times6\times6$ (pictured in insets).
    $D$, $\chi$, the algorithm and its hyper-parameters are all uniformly sampled.
    }
    \label{fig:trees-cost-vs-mem}
\end{figure}

There are various cost functions one can assign to an approximate contraction tree to then optimize against.
Broadly these correspond to either space (memory) or time (FLOPs) estimates.
Three cost functions that we have considered that only depend on the tree and $\chi$ (but not gauging scheme for example) are the estimated peak memory, $M$, the largest intermediate tensor, $W$, and the number of FLOPs involved in the contractions only.
Specifically, given the set of tensors, $\{v_\alpha\}$, present at stage $\alpha$ of the contraction, the peak memory is given by:
\begin{equation}
    M = \max_{\alpha} \sum_{v \in \{v_\alpha\}} \mathrm{size}(T^{[v]})~.
\end{equation}

Given a compression and gauging scheme, one can also trace through the full computation, yielding a more accurate peak memory usage, $\tilde{M}$, as well an estimate of the FLOPs associated with all QR and SVD decompositions too -- we call this the full computational `cost', $C$. Included in this we consider only the dominant contributions:
\begin{itemize}
    \item contraction of two tensors with effective dimensions $(m, n)$ and $(n, k)$: $mnk$
    \item QR of tensor with effective dimensions $(m, n)$ with $m\geq n$: $2 m n^2 - \frac{2}{3} n^3$
    \item SVD of tensor with effective dimensions $(m, n)$ with $m\geq n$: $4 m n^2 - \frac{4}{3} n^3$.
\end{itemize}
Of these the first two dominate since the SVD is only ever performed on the reduced bond matrix. Note the actual FLOPs will be a constant factor higher depending on the data type of the tensors.

In Fig.~\ref{fig:trees-cost-vs-mem} we plot the relationship between the various metrics mentioned above for several thousand randomly sampled contraction trees on both a square and cubic geometry for varying $D$, $\chi$ and algorithm.
We note that $M$, $\tilde{M}$ and $W$ are all tightly correlated.
The full cost $C$ is slightly less correlated with these and only slightly more so with the `contractions only` cost.
Importantly however, the best contractions largely appear to simultaneously minimize all the metrics.

\section{Hand-coded Contraction Schemes}

\begin{figure}
    \centering
    \includegraphics[width=\linewidth]{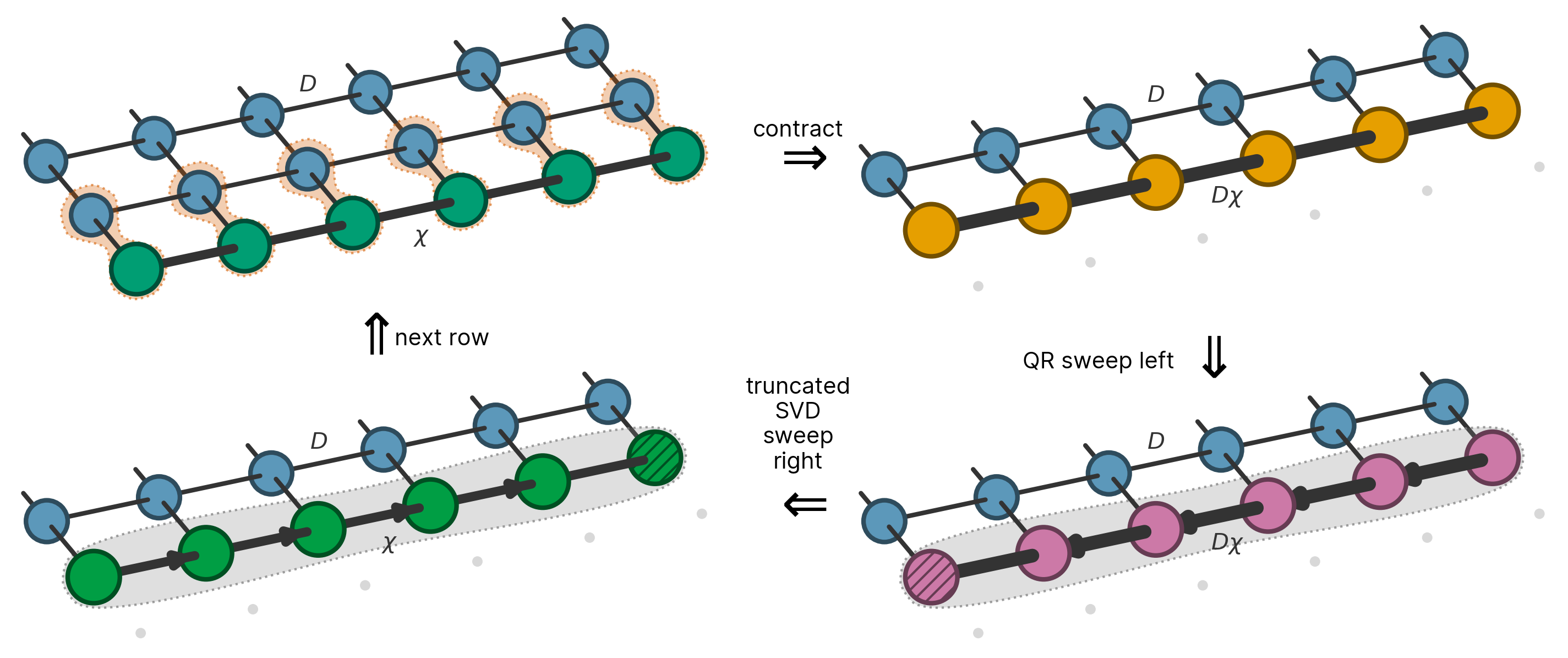}
    \caption{Overview of a single step of the the manual 2D boundary contraction method
    that uses an MPS to sweep across the square.
    }
    \label{fig:contract_2d_sweep}
\end{figure}

\begin{figure}
    \centering
    \includegraphics[width=\linewidth]{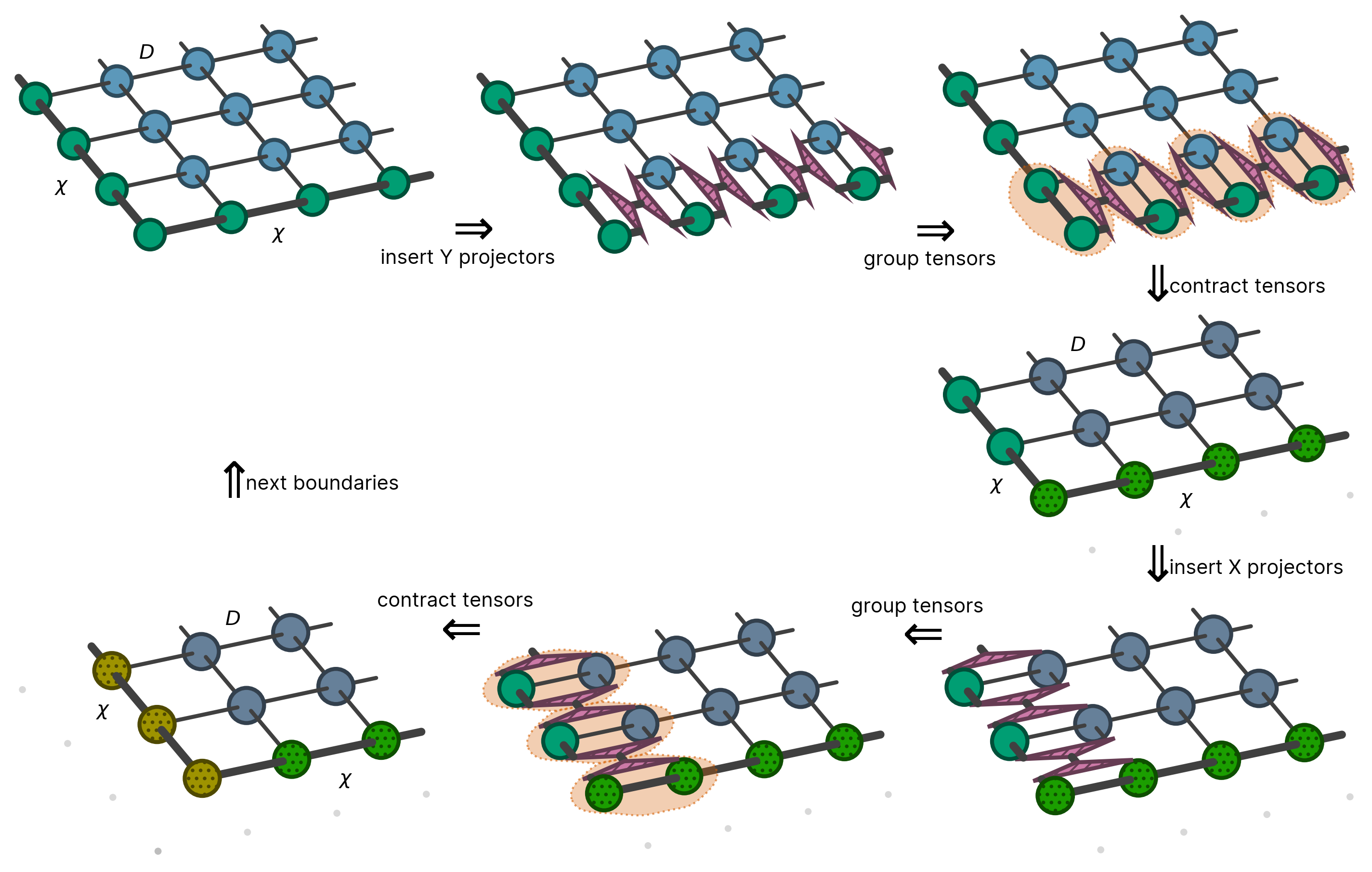}
    \caption{
    Illustration of two boundary contraction steps of CTMRG for a finite 2D lattice. The full algorithm proceeds to contract all four of the sides inwards in succession.
    Note that the projectors (pink) are not identical across the lattice but are computed specific to the local tensors to allow for finite in-homogeneous systems.
    }
    \label{fig:contract_2d_ctmrg}
\end{figure}

\begin{figure}
    \centering
    \includegraphics[width=\linewidth]{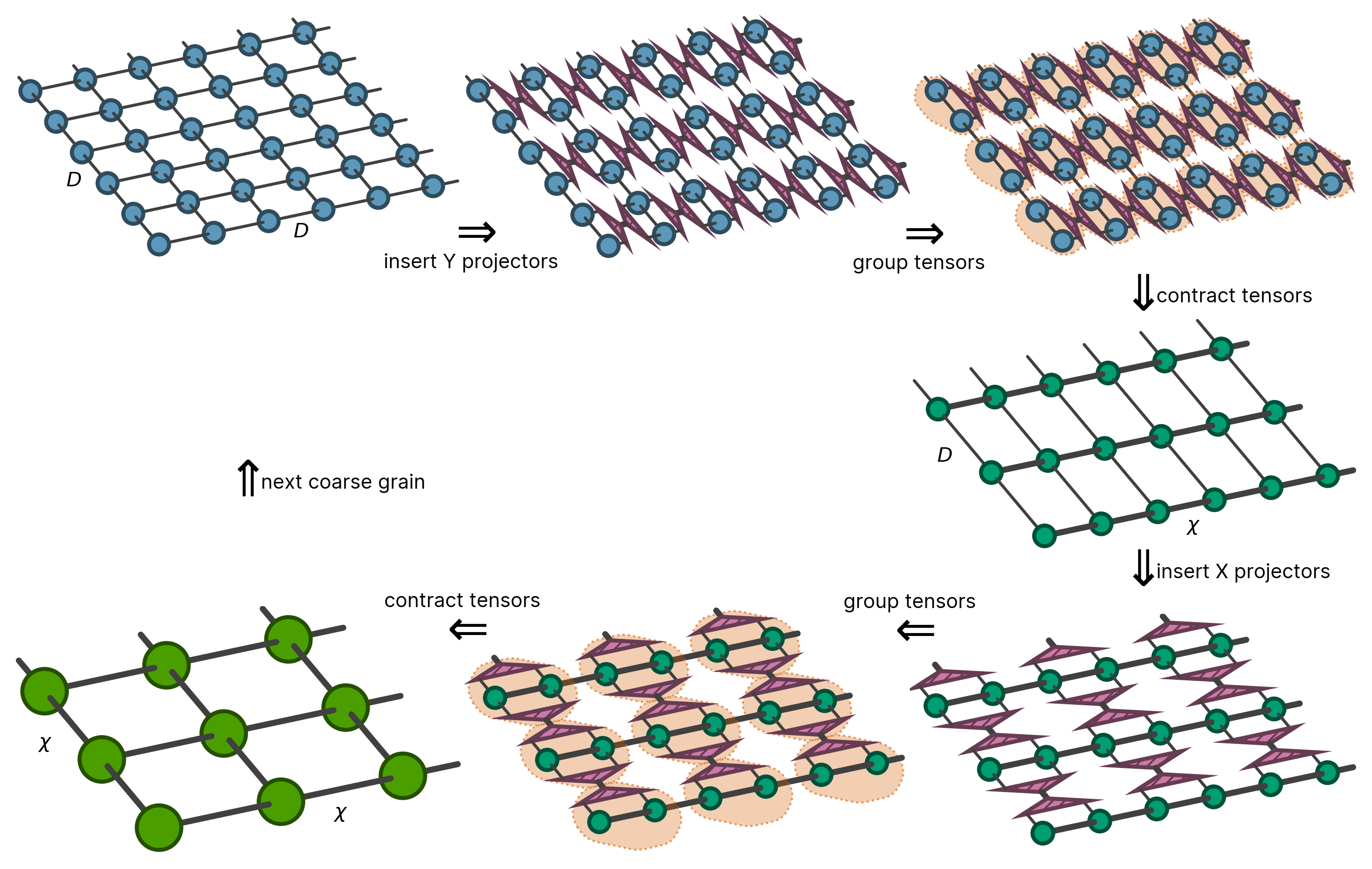}
    \caption{
    Illustration of a full coarse graining step of HOTRG for a finite 2D lattice.
    Note that once a round of coarse graining has taken place, all bonds are of size $\chi$ and so the next round starts with $D=\chi$.
    Note also that the projectors (pink) are not identical across the lattice but are computed specific to the local tensors to allow for finite in-homogeneous systems.
    }\label{fig:contract_2d_hotrg}
\end{figure}

\begin{figure}
    \centering
    \includegraphics[width=\linewidth]{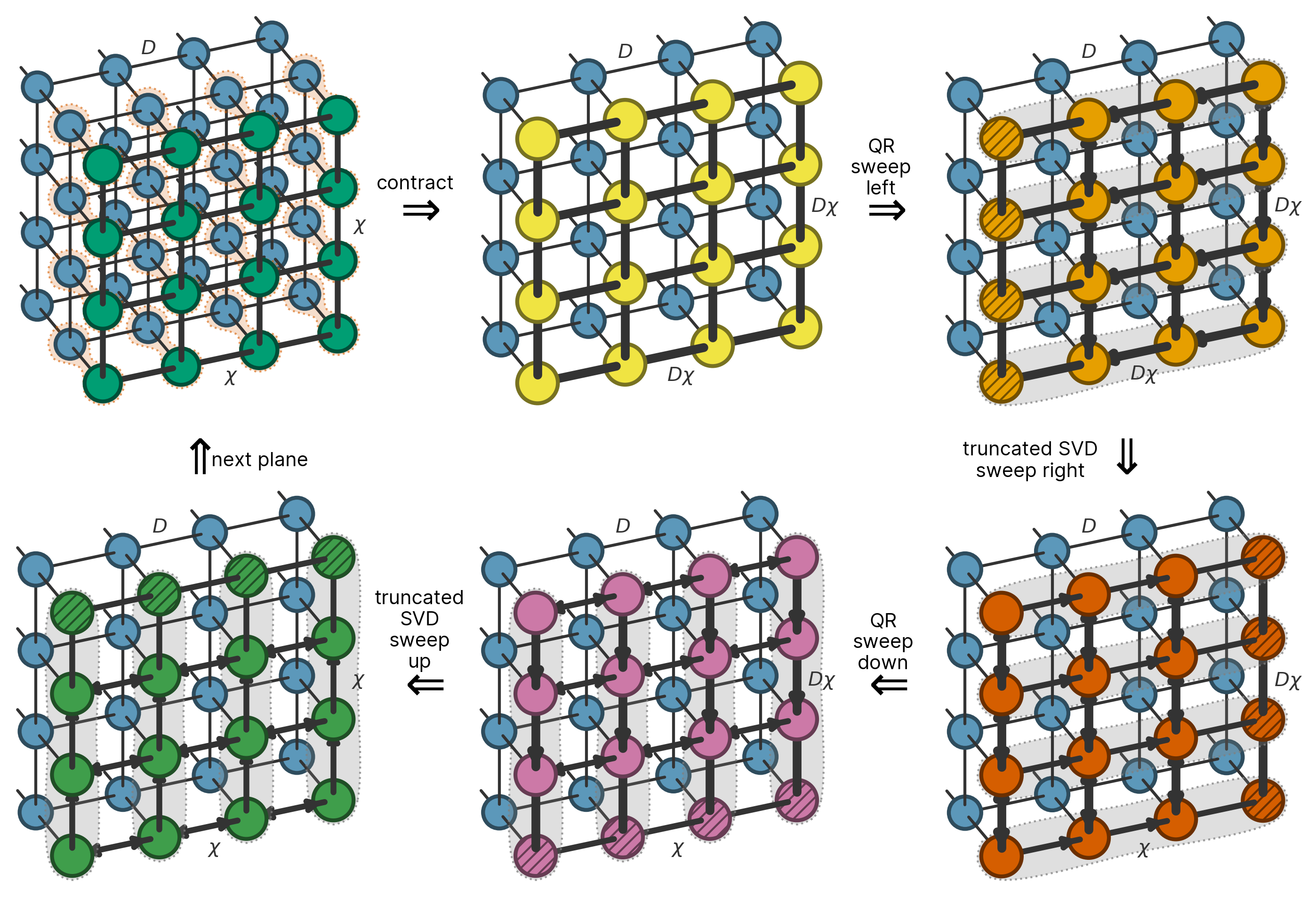}
    \caption{Illustration of a single step of the the manual 3D boundary contraction method
    that uses a PEPS to sweep across the cube. When $L_x$, $L_y$ or $L_z = 1$ the scheme becomes equivalent to MPS boundary contraction.
    }\label{fig:contract_3d_sweep}
\end{figure}

\begin{figure}
    \centering
    \includegraphics[width=0.75\linewidth]{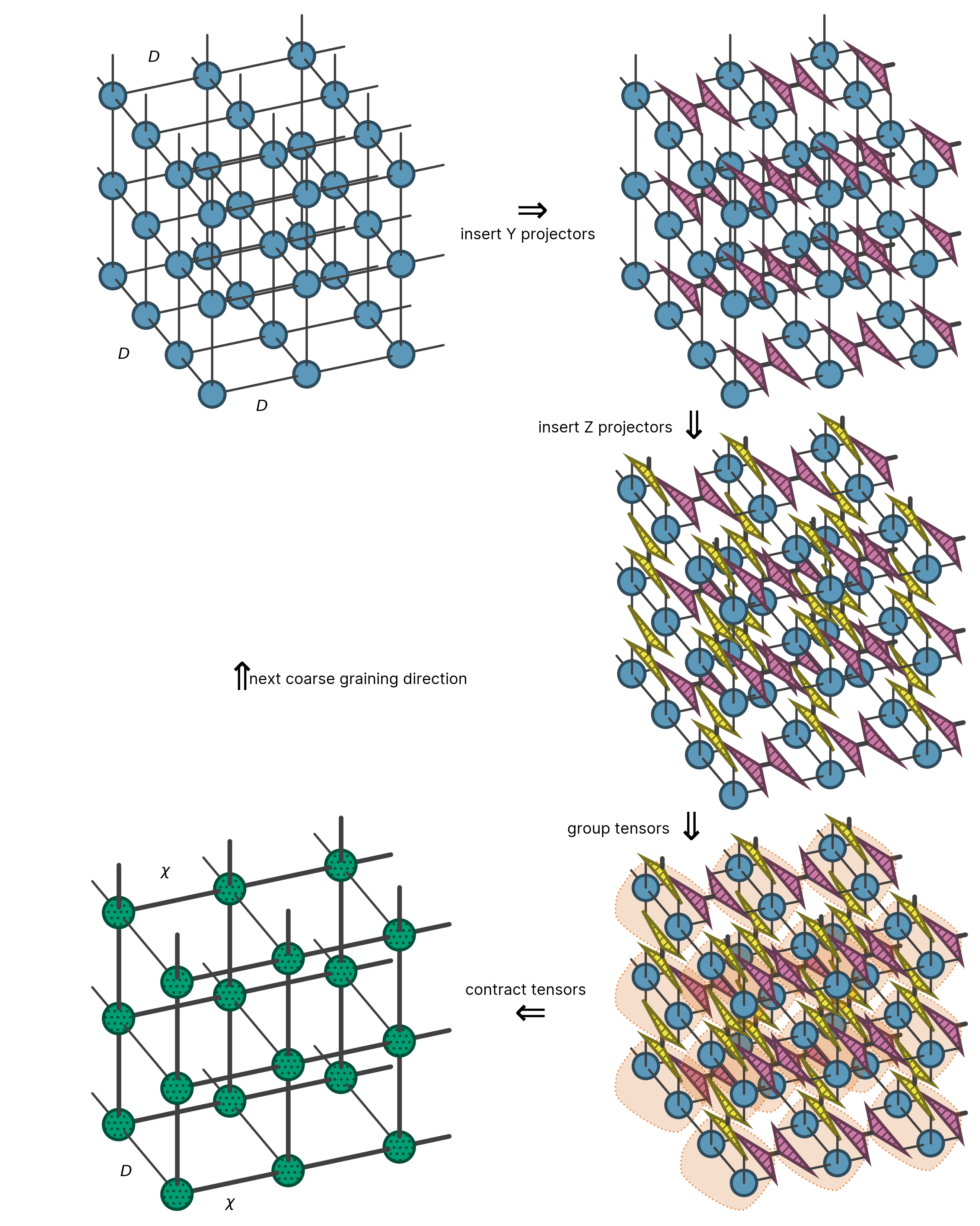}
    \caption{
        Illustration of a single coarse graining step of HOTRG for a finite 3D lattice. For brevity we only show coarse graining in the x-direction but the full algorithm coarse grains each of the three dimensions in succession. Note that once two or more directions have been coarse grained, all bonds will be of size $\chi$ and so subsequent rounds start with $D=\chi$.
        Note also that the projectors (pink and yellow) are not identical across the lattice but are computed specific to the local tensors to allow for finite in-homogeneous systems.
    }\label{fig:contract_3d_hotrg}
\end{figure}

In Figs.~\ref{fig:contract_2d_sweep}-\ref{fig:contract_3d_hotrg} we illustrate the various hand-coded contraction schemes used as comparisons in the text: 2D boundary contraction, 2D corner transfer matrix RG~\cite{nishino1996corner}, 2D higher-order TRG~\cite{xie2012coarse}, 3D PEPS boundary contraction, and 3D higher-order TRG~\cite{xie2012coarse}. Note that in the case of CTMRG and HOTRG, the algorithms are usually iterated to treat infinite, translationally invariant lattices, but here we simply apply a finite number of CTMRG or HOTRG steps and also generate the projectors locally to handle in-homogeneous tensor networks.
For both CTMRG and HTORG we use the cheaper, `lazy' method~\cite{iinoBoundaryTensorRenormalization2019} of computing the reduced factors $R_A$ and $R_B$ which avoids needing to form and compute a QR on each pair of tensors on either side of a plaquette. We then use the projector form as given in Sec.~\ref{sec:projectors} to compress the plaquette.
The 3D PEPS boundary contraction algorithm has not previously been implemented to our knowledge, but is formulated in a way analogous to 2D boundary contraction. Notably, if any dimension is of size 1 it reduces to exactly 2D boundary contraction including canonicalization. For further details, we refer to the lecture notes~\cite{ranLectureNotesTensor2020} and the original references.

\section{Models}

\subsection{Ising Model}

We consider computing the free energy per spin of a system of $N$ classical spins at inverse temperature $\beta$,
\begin{equation}
    f = \frac{-\log Z}{N \beta}
\end{equation}
where the partition function, $Z$, is given by:
\begin{equation}\label{eq:ising-partition-fn}
    Z = \sum_{\{\sigma\}} \prod_{\langle i, j \rangle} \exp(j \beta \sigma_i \sigma_j)~,
\end{equation}
$\sigma_k \in [1, -1]$ being the state of spin $k$ and $\{\sigma\}$ the set of all configurations. The interaction pairs $\langle i, j \rangle$ are the edges of the graph, $G$, under study. We take the interaction strength $j$ to be 1, i.e. ferromagnetic.
While Monte Carlo methods can readily compute many quantities in such models, we note that the partition function and free energy are typically much more challenging~\cite{wangEfficientMultipleRangeRandom2001}.
Regardless of geometry we assume the spins are orientated in the same direction - the uniaxial Ising model.
Typically one converts $Z$ into a `standard' tensor network with a single tensor per spin (or equivalently vertex of $G$), by placing the tensor,
\begin{equation}
    T^{[v]}_{\{e_v\}} = \sum_{i} \prod_{e_j \in \{e_v\}} W_{i, e_j}
\end{equation}
on each vertex $v$ of $G$, where the matrix $W$ is defined by $(W^2)_{i,j} = \exp(\beta \sigma_i \sigma_j)$.
For $j>0$ we can define $W$ as real and symmetric using:
\[
W =
\dfrac{1}{\sqrt{2}}
\begin{pmatrix}
\sqrt{\cosh(j \beta)} {+} \sqrt{\sinh(j \beta)} &
\sqrt{\cosh(j \beta)} {-} \sqrt{\sinh(j \beta)} \\
\sqrt{\cosh(j \beta)} {-} \sqrt{\sinh(j \beta)} &
\sqrt{\cosh(j \beta)} {+} \sqrt{\sinh(j \beta)}
\end{pmatrix}~.
\]
This is equivalent to splitting the matrix on each bond then contracting each factor into a COPY-tensor placed on each vertex.
We note that while this yields a tensor network with the exact geometry of the interaction graph $G$, one could factorize the COPY-tensor in other low-rank ways.
Indeed for the exact reference results the TN in Eq.~\eqref{eq:ising-partition-fn} is contracted directly by interpreting every spin state index as a hyper index (i.e. appearing on an arbitrary number of tensors). The relative error in the free energy is given by:
\begin{equation}
    \Delta f
    = \left|1 - \frac{f}{f_{\mathrm{exact}}}\right|
    = \left|1 - \frac{\log Z}{\log Z_{\mathrm{exact}}}\right|
\end{equation}
with $\cdot_{\mathrm{exact}}$ results obtained via exact contraction.
Depending on geometry the Ising model undergoes a phase transition at critical temperature $\beta_c$ in the thermodynamic limit and it is in this vicinity that generally $\Delta f$ peaks for finite systems.
For example, on the 2D square lattice the exact value is known, $\beta_c=\frac{\log(1 + \sqrt{2})}{2} \approx 0.44$~\cite{onsager1944crystal,baxter2016exactly}.

\subsection{URand Model}

While the Ising model varies in difficulty depending on $\beta$, it seems always relatively easy to approximate to some extent using approximate contraction.
On the other hand we expect there to be tensor networks which are exponentially difficult to approximate even for simple geometries.
Here we introduce the \emph{URand} model which allows us to continuously tune between very hard and very easy regimes.
This is achieved simply by filling each tensor with random values sampled uniformly from the range $[\lambda, 1]$.
When $\lambda \geq 0$, every term in the TN sum is non-negative and the sum becomes very easy to approximate.
As $\lambda$ becomes more negative however, the sum increasingly becomes terms of opposite sign which `destructively interfere' making the overall contracted value $Z$ hard to approximate.
Choosing an intermediate $\lambda$ allows us to generate `moderately hard' contractions where the different gauging and tree generating strategies have a significant effect.
For the URand model we consider the relative error in $Z$ directly:
\[
\Delta Z = 1 - \dfrac{Z}{Z_\mathrm{exact}}
\]
with $Z_\mathrm{exact}$ computed via exact contraction.

\subsection{Dimer covers / positive \#\textsc{1-in-3SAT}}

In this model we want to compute the entropy per site of dimer coverings of a graph $G$ with number of vertices $|V|$.
Here, a valid configuration is given if every vertex of $G$ is `covered' by exactly one dimer.
Counting all valid configurations is done by enumerating every combination of placing a dimer on a bond (setting the corresponding index to 1) or not (setting the index to 0), which can be formed as a tensor network with the following tensor on each vertex:
\begin{equation}
    T_{i, j, k, \ldots} =
    \begin{cases}
    1, & \text{if} \ i + j + k + \ldots = 1 \\
    0, & \text{otherwise}
    \end{cases}~.
\end{equation}
The total number of valid configurations is then the contraction:
\begin{equation}
    W = \sum_{\{e\}} \prod_{v} T^{[v]}_{\{e_v\}}
\end{equation}

This is also equivalent the counting problem positive \#\textsc{1-in-3SAT}~\cite{raymondPhaseDiagram1in32007,zdeborovaConstraintSatisfactionProblems2008,kourtis2019fast}, the decision version of which is NP-Complete~\cite{schaeferComplexitySatisfiabilityProblems1978,gareyComputersIntractabilityGuide1979}.
For 3-regular random graphs, this is known to be close the hardest regime though just on the side of satisfiability, in terms of the density of variables (edges) to clauses (vertices)~\cite{raymondPhaseDiagram1in32007,zdeborovaConstraintSatisfactionProblems2008}.
The residual entropy per site is given by:
\begin{equation}
    S = \frac{\log W}{n}~,
\end{equation}
for number of vertices $n=|V|$, with relative error:
\begin{equation}
    \Delta S = 1 - \dfrac{S}{S_\mathrm{exact}}~.
\end{equation}
The reference values $S_\mathrm{exact}$ are computed using exact contraction, which is feasible up to $n \sim 300$ for 3-regular random graphs.

The problem is also known as counting `perfect matchings', `complete matchings', or `1-factors' and has been studied for random regular graphs in the large $|V|$ limit~\cite{bollobas1986number}.
There it was shown that if the degree $k$ satisfies $3 \leq k < \log(n)^{\frac{1}{3}}$ then the expected value of $W$ across all random $k$-regular instances is
\begin{align}
\bar{W} =
(\sqrt{2} + O(n^{-\frac{2}{3}}))
e^{\frac{1}{4}} ((k - 1)^{k - 1}/k^{k - 2} )^{ n/2 }~.
\end{align}
If we take the limit of this we find:
\begin{align*}
s_\infty &= \lim_{n\rightarrow \infty} \left( \log \hat W / n \right)\\
         &= \lim_{n\rightarrow \infty} \left( \dfrac{1}{2} (k - 1) \log(k - 1) + \dfrac{1}{2} (2 - k) \log(k) + O(\dfrac{1}{n})   \right) \\
         &= 0.1438410362258904\ldots
\end{align*}
The condition linking $k$ and $n$ requires $n \gtrsim 5.3 \times 10^{11}$, the scale of which suggests that our estimate of $0.1429(2)$ might have some small systematic error remaining from finite size effects.

\section{Performance comparison to \texttt{CATN}}

So far where appropriate we have compared our method to manually specified contraction orders in 2D and 3D.
In~\cite{pan2020contracting}, an algorithm to automatically contract arbitrary geometry tensor networks was also developed which showed good performance across a range of graphs.
For convenience here we refer to that algorithm as \texttt{CATN}.
While the basic tensor operations are similar, in \texttt{CATN} an effective periodic MPS is used to contract the graph using SWAP operations to remove bonds.
A major difference is also that in this work we optimize the pattern of contractions and compressions ahead of time for the specific geometry.
A python implementation of~\cite{pan2020contracting} is available at~\cite{catn} which we can use to compare against for free energies of graphical models.
Taking that code as is, the most direct way to compare performance of these two approaches is simply wall time on a single core of a CPU, here an AMD EPYC 7742.

\begin{figure}[t!]
    \centering
    \includegraphics[width=\textwidth]{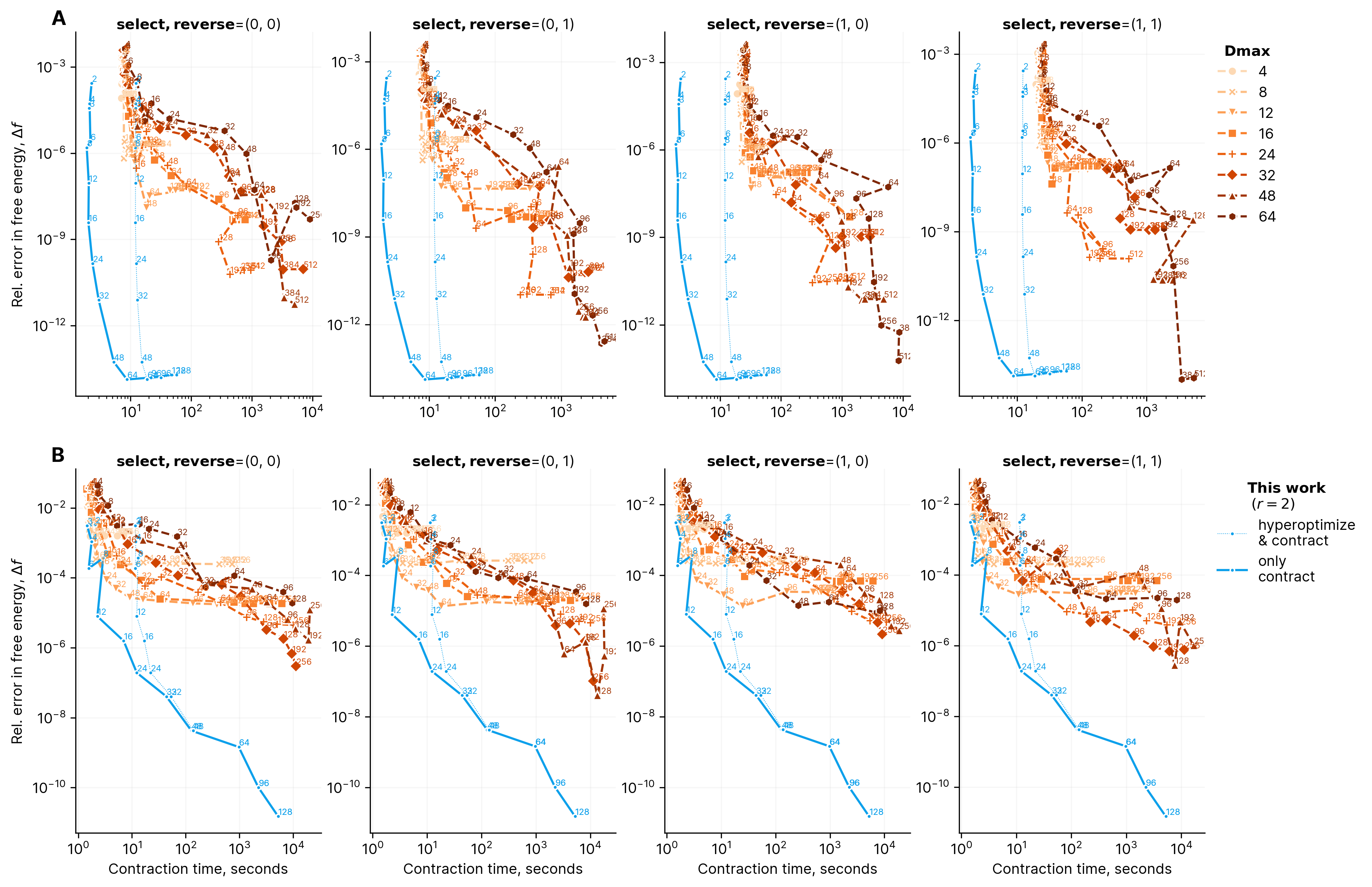}
    \caption{
        Performance comparison of this current work (blue) and the algorithm of Pan et al.~\cite{pan2020contracting} (orange) using various settings for computing the free energy of the Ising model at approximately the critical point of \textbf{A}: a $32{\times}32$ square lattice at the approximate critical temperature $\beta{=}0.44$ and \textbf{B}: a $6{\times}6{\times}6$ cubic lattice at the approximate critical temperature $\beta{=}0.3$.
        For both algorithms $\chi$ is varied and the points are labeled with the value.
        For the current work we show both the time with contraction only, and also accounting for the hyper-optimization time (about 10 seconds).
    }
    \label{fig:detailed-catn-comparison}
\end{figure}

In Fig.~\ref{fig:detailed-catn-comparison} we show a more detailed comparison than the main text of \texttt{CATN} and this current work for the Ising model at approximately the critical point on 2D and 3D lattices, as a function of accuracy vs contraction time.
For our algorithm here we use the \texttt{Span} tree builder and show a range of $\chi$ with the tree gauge distance $r=2$.
One consideration is that in our approach the hyper-optimization step might run separately to the actual contraction, since it depends only on the geometry and the approximate contraction tree can be re-used for different tensor entries (e.g. sweeping $\beta$).
Here we show both the pure contraction only time and also the time if one takes the hyper-optimization into account.
We compare to the contraction time reported directly by \texttt{CATN}, and note that this includes computing (greedily) which edge to remove next on-the-fly.
In \texttt{CATN}, there are two bond dimension parameters controlling the trade-off between accuracy and computational effort, $D_\mathrm{max}$ and $\chi$, and two main parameters controlling how to select the bonds to remove, \texttt{select} and \texttt{reverse}.
The relationship between $D_\mathrm{max}$ and $\chi$ and the error and computational time is not trivial so we sweep across both.
For the main text we showed \texttt{select=0} and \texttt{reverse=1} but here we also show the other good combinations.
\texttt{CATN} also has four other parameters, \texttt{node}, \texttt{corder}, \texttt{swapopt}, and \texttt{svdopt}.
We find these generally have no systematic or significant effect on error time for these examples, but nonetheless for each point take the best performing combination in terms of $\min \Delta F \times \mathrm{time}$.
The core linear algebra operations in both algorithms are performed using the same version of \texttt{numpy}~\cite{harris2020array}, and we take the best time out of three repeats.

In both the 2D and 3D cases, Figs.~\ref{fig:detailed-catn-comparison}A and B respectively, we see that our algorithm achieves the best accuracy vs. contraction time trade-off across the range of values and settings considered, especially in the high accuracy regime.
We note that once hyper-optimization times are taken into account, for some less high accuracies \texttt{CATN} can perform better.
For this comparison, we are interested in the automated performance of each algorithm, and thus use the greedy ordering of \texttt{CATN}. However, it was noted in~\cite{pan2020contracting} that an explicitly specified `Zig-Zag' order performs well for the square 2D lattice, suggesting that optimizing over strategies might be beneficial for that approximate contraction scheme also.

\section{Performance on corner double line (CDL) tensor networks}

\begin{figure}[t!]
    \centering
    \includegraphics[width=0.6\linewidth]{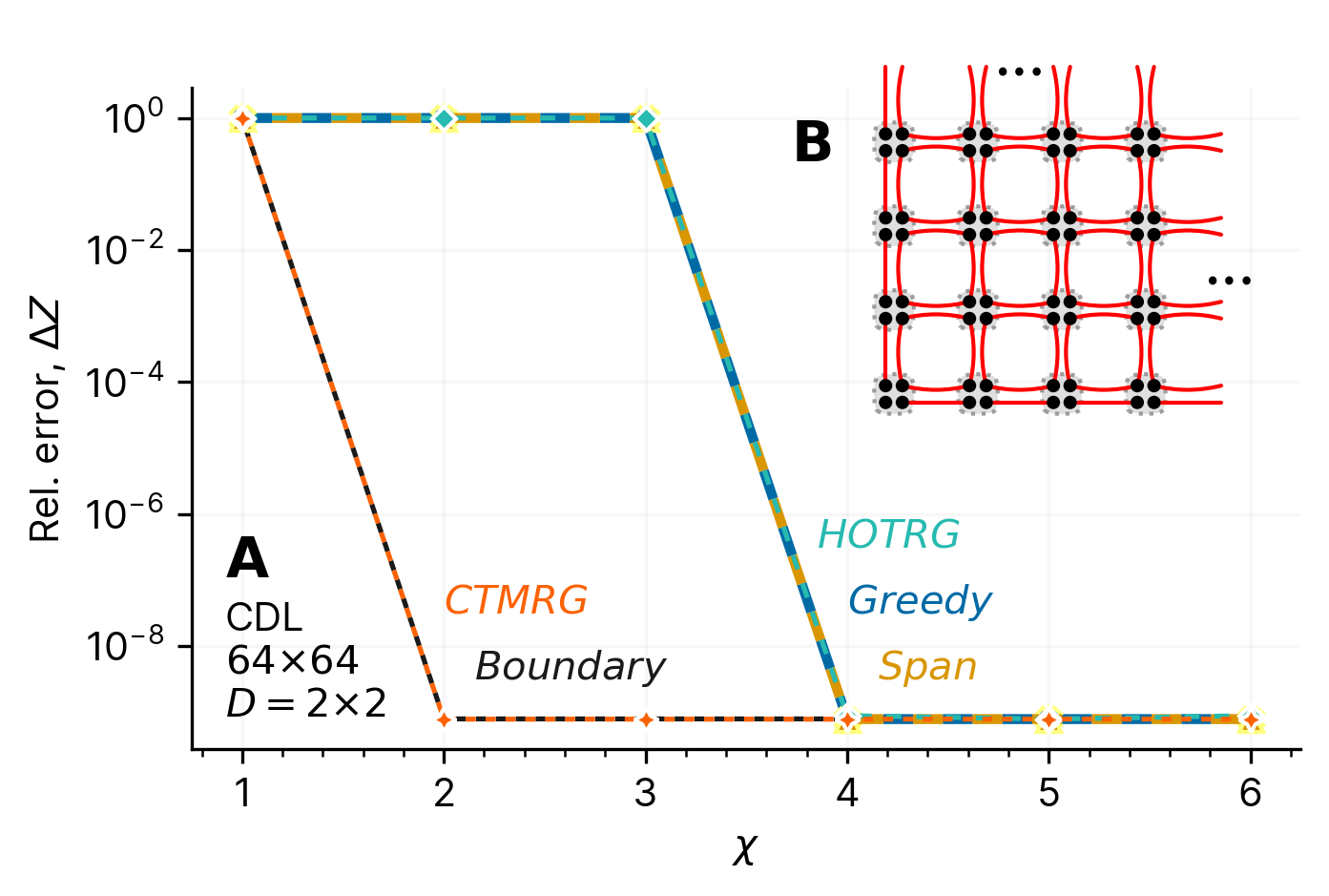}
    \caption{
    \textbf{A}: Accuracy of both manual and hyper-optimized approximate contraction schemes for the 2D corner double line (CDL) model with lattice size of $64\times64$ and an effective $D=2\times2=4$.
    \textbf{B}: schematic of the OBC CDL model.
    }
    \label{fig:cdl-compare}
\end{figure}

An important model in the development of various normalization group style approximate contraction algorithms has been that of the corner double line (CDL) TN~\cite{evenblyTensorNetworkRenormalization2015}.
This model involves embedding local loop correlations in a lattice. Each such loop consists of four 2 dimensional COPY tensors (i.e. identity matrices) with dimension $d$ placed around each \textit{plaquette}. This results in four corner tensors at each \textit{site} which can be contracted (via an outer product) to give a $d^2 \times d^2 \times d^2 \times d^2$ tensor after grouping indices. Each bond is thus doubled, carrying correlations from the two adjacent plaquettes.
Such a CDL TN has a trivial, purely local correlation structure, that should not propagate to the coarse grained picture after a real space normalization procedure.
However it is simple to show that both elementary algorithms such as TRG~\cite{levin2007tensor} and also more advanced algorithms such as HOTRG~\cite{xie2012coarse} never fully remove all such correlations, and it has been speculated that this is a source of error in approximate contraction schemes for more physically motivated models, which has sparked many improved schemes that explicitly handle the CDL correlations.

On the other hand, if one is only interested in the accuracy of the contracted value of the TN, then the CDL model poses no problem for all the contraction methods we consider here. Indeed the correlations are exactly the type that can be sustainably removed as the contraction proceeds, as long as $\chi$ is above some very small threshold.
In Fig.~\ref{fig:cdl-compare}A we compare the relative error for contracting a $64{{\times}}64$ CDL TN with 5 methods and show that each becomes essentially exact at either $\chi=2$ or $\chi=4$ (for $d=2$).
Since we focus on open boundary conditions here, we use the CDL TN depicted in Fig.~\ref{fig:cdl-compare}B with 0- and 1- dimensional COPY tensors along the boundary where necessary, however the same behavior holds for periodic boundary conditions.
This is easily understood from the fact that once two pairs of tensors on either side of a plaquette have been contracted to $A, B$ (which all these methods do) the internal plaquette correlation is `resolved' into a scalar contribution, allowing the remaining local operator $AB$ to be exactly represented with rank reduced by a factor of $d^2$.

\end{document}